\documentclass[11pt,a4paper]{article}
\pdfoutput=1
\usepackage{jcappub}
\usepackage{rotating}
\usepackage{array}
\usepackage{amsmath}
\usepackage[normalem]{ulem}
\usepackage{slashed}
\usepackage{placeins}
\usepackage{graphicx}
\usepackage{geometry,amsmath,amsfonts}
\usepackage{epsfig}
\usepackage[normalem]{ulem}
\usepackage{dcolumn}
\usepackage{bm}
\usepackage{amssymb}
\usepackage{color}
\usepackage{subfig}
\def\gsim{\lower0.5ex\hbox{$\:\buildrel >\over\sim\:$}}
\def\lsim{\lower0.5ex\hbox{$\:\buildrel <\over\sim\:$}}

\newcommand{\be}{\begin{equation}}
\newcommand{\ee}{\end{equation}}
\newcommand{\bea}{\begin{eqnarray}}
\newcommand{\eea}{\end{eqnarray}}
\newcommand{\ben}{\begin{enumerate}}
\newcommand{\een}{\end{enumerate}}
\newcommand{\bde}{\begin{widetext}}
\newcommand{\ede}{\end{widetext}}

\newcommand{\bc}{\begin{center}}
\newcommand{\ec}{\end{center}}

\begin{document}

\title{Pseudoscalar Mediators: A WIMP model at the Neutrino Floor}
\author{Giorgio Arcadi$^1$}
\emailAdd{arcadi@mpi-hd.mpg.de}
\author{Manfred Lindner$^1$}
\emailAdd{lindner@mpi-hd.mpg.de}
\author{Farinaldo S. Queiroz$^{1,2}$}
\emailAdd{queiroz@mpi-hd.mpg.de}
\author{Werner Rodejohann$^1$}
\emailAdd{werner.rodejohann@mpi-hd.mpg.de}
\author{Stefan Vogl$^1$}
\emailAdd{stefan.vogl@mpi-hd.mpg.de}

\affiliation{$^1$Max\--Planck\--Institut f\"ur Kernphysik (MPIK), Saupfercheckweg 1, 69117 Heidelberg, Germany}
\affiliation{$^2$International Institute of Physics, Federal University of Rio Grande do Norte, Campus Universit\'ario,
Lagoa Nova, Natal-RN 59078-970, Brazil}

\abstract{
Due to its highly suppressed cross section (fermionic) dark matter interacting with the Standard Model via pseudoscalar mediators is expected to be essentially unobservable in direct detection experiments. We consider both a simplified model and a more realistic model based on an extended two Higgs doublet model and compute the leading one-loop contribution to the effective dark matter-nucleon interaction. This higher order correction dominates the scattering rate completely and can naturally, i.e. for couplings of order one, lead to a direct detection cross section in the vicinity of the neutrino floor. Taking the observed relic density and constraints from low-energy observables into account we analyze the direct detection prospects in detail and find regions of parameter space that are within reach of upcoming direct detection experiments such as XENONnT, LZ, and DARWIN.}

\maketitle

\section*{Introduction}%

The presence of dark matter (DM) in our universe has been established in a variety of datasets~\cite{Bertone:2010zza} and the observations of the Planck satellite show that it accounts for 27\% of the energy content of the Universe~\cite{Ade:2015xua}. One of the most compelling dark matter candidates is a  massive particles with weak-scale interactions, a so-called WIMP (Weakly Interacting Massive Particle)~\cite{Arcadi:2017kky}. Thermal freeze-out of WIMPs furnishes a compelling solution for the DM puzzle since it correlates elegantly the DM relic density  with the DM interaction strength with Standard Model (SM) particles via a single particle physics input, i.e.\ the thermally averaged pair annihilation cross section. 
In many scenarios this annihilation cross section is tied, by crossing symmetry, to the scattering cross section on nucleons; the experimentally favored value of the former frequently corresponds to a value of the latter that is in conflict with null results from direct DM searches~\cite{Agnese:2014aze,Agnes:2015ftt,Angloher:2015ewa,Akerib:2016vxi,Hehn:2016nll,Undagoitia:2015gya,Tan:2016zwf,Amole:2017dex,Aprile:2017iyp,Amaudruz:2017ekt}. This correlation can be weakened by considering next-to-minimal scenarios (see e.g.~\cite{Duerr:2016tmh,Arcadi:2016qoz,Arcadi:2017vis}) or models that lead to suppressed scattering scatterings rates due to the low energy scale in the process~\cite{Berlin:2014tja,Arina:2014yna,Dolan:2014ska,Abdullah:2014lla,Hektor:2014kga,No:2015xqa,Liu:2015oaa,Fan:2015sza,Buchmueller:2015eea,Hektor:2015zba,Berlin:2015wwa,Karwin:2016tsw,Banerjee:2017wxi,Bauer:2017ota,Baek:2017vzd,Hektor:2017ftg} such as the one investigated here.

The latter takes advantage of the different energy scales involved in the annihilation and scattering process to loosen the correlation. The scale for the annihilation cross section is set by the dark matter mass, whereas the momentum transfer in the dark matter-nucleon scattering is only $\mathcal{O}(100)$ keV. This can be exploited in the case of fermionic DM interacting via a light pseudoscalar field. As pointed out in~\cite{Freytsis:2010ne,Boehm:2014hva} the DM scattering cross section on nucleons at tree level is proportional to the fourth power of the momentum transfer and, therefore, the scattering rate in realistic direct detection experiments is essentially negligible (see however~\cite{Arina:2014yna,Yang:2016wrl}).  The DM pair annihilation cross section, on the other hand, does not suffer from any suppression and the observed relic density can easily be generated by thermal freeze-out. 

In the following, we investigate whether a non-negligible DM scattering rate can be generated at higher order in such a model and compute the DM nucleus scattering cross section. It turns out that for typical values of the involved couplings and masses, the loop-induced direct detection cross section takes values in the vicinity of the so-called ``neutrino floor''~\cite{Billard:2013qya}, i.e.\ the cross section corresponding to the coherent scattering of neutrinos on nucleons. This process will induce  a signal which is similar to the elastic scattering of a WIMP and thus represents an irreducible background~\cite{Ruppin:2014bra,Davis:2014ama,Dutta:2015vwa,Dent:2016iht,Ng:2017aur}. Despite possibilities of discriminating signals from WIMP and neutrino scattering, for example by combining detectors with different target materials, the neutrino floor is customarily regarded as the ultimate sensitivity for future Direct Detection experiments such as XENONnT~\cite{Aprile:2015uzo}, LZ~\cite{Szydagis:2016few} and DARWIN~\cite{Aalbers:2016jon}.

 The strength of a  direct detection signal arising from high order corrections should be compared with existing constraints from low energy probes and collider searches to assess the importance of these loop effects in the WIMP-nucleon scattering. Throughout our analysis we assume that the entire dark matter relic density is determined  by thermal freeze-out and do consider modification which could arise in non-standard cosmologies.

A reliable comparison between different observables requires going beyond a simplified setup. For this reason, we  investigate the dark matter phenomenology in a simplified DM model as well as in a full UV-complete model~\cite{Goncalves:2016iyg,Bauer:2017ota}. We  show that the DM phenomenology of the simplified model differs from the full model due to the presence of new particles that cannot be fully decoupled.

The structure of this paper is as follows: First, we introduce a minimal simplified model for fermionic dark matter interacting with the SM via a pseudoscalar mediator. We discuss the phenomenology of the model with a  particular emphasis on observables that are relevant for light pseudoscalars. In a second step, we generalize the simplified model and embed it in a gauge-invariant, UV-complete model. We investigate whether the conclusions derived in the simplified model persist in the more general framework and comment on additional observables which become relevant in this case before concluding. 
 
\section{Simplified model}%
\label{sec:Simple} 

The model under consideration consists of a Dirac fermion $\chi $~\footnote{The case of Majorana DM is qualitatively the same but minor quantitative differences arise since DM is its own antiparticle in this case.} which is a singlet under the SM gauge group and acts as our DM candidate. The interactions of $\chi$ with the SM are mediated by a $s$-channel pseudoscalar mediator $a$ and can be described by the Lagrangian:
\begin{equation}
\label{eq:tree_lag}
\mathcal{L}=i a \left( g_\chi \bar{\chi}\gamma_5 \chi+ c_a \sum_f \frac{ m_f}{v_h}\bar{f}\gamma_5 f \right),
\end{equation}  
where $f$ is a SM fermion and $v_h=246\,\mbox{GeV}$ the vacuum expectation value of the SM Higgs.
\noindent
We have assumed Yukawa-like couplings of the pseudoscalar with the SM fermions and parameterized our ignorance regarding the origin of this couplings by the rescaling parameter $c_a$ (we will discuss a more concrete realization in the next section), while we have been agnostic concerning the DM coupling $g_\chi$. 
The simplified model defined by this Lagrangian has only 4 free parameters, i.e.\ the masses $m_\chi$ and $m_a$ of the new particles, the DM coupling $g_\chi$ and the rescaling parameter $c_a$.

As pointed out previously, the main goal of this work is to scrutinize potential direct detection prospects and we will focus our attention on the promising regions of parameter space.
We consider masses of the pseudoscalar in the range $1 \leq m_a \leq 100\,\mbox{GeV}$. The lower limit of the mass is chosen since it corresponds on the typical energy scales of nuclear processes. For masses of the mediator below this value, the conventional treatment of DM direct detection becomes questionable; we leave this to further study. As will be shown in the following, direct detection is irrelevant for $m_a \gtrsim 100\,\mbox{GeV}$. 
Heavier mediator masses can be constrained by LHC searches for decays of $a$ to  SM fermions~\cite{Banerjee:2017wxi} while searches which target the invisible decay of $a$ are currently not competitive~\cite{Buckley:2014fba,Harris:2014hga}.
Since direct detection loses sensitivity for light dark matter we limit our study to $m_\chi \gtrsim 10$ GeV in what follows.

\subsection{Dark Matter Annihilations and the Relic Density}

Provided that the dark matter has been in thermal equilibrium with the SM plasma in the early Universe its present relic density is set by the abundance at freeze-out. Hence the dark matter density $\Omega_{\chi} h^2$  is determined by the thermally averaged pair annihilation cross section $\langle \sigma v \rangle$ of the DM. 
The observed value of $\Omega_{\chi} h^2 \simeq 0.12$~\cite{Ade:2015xua} is achieved for 
$\langle \sigma v \rangle \simeq 3 \times 10^{-26}\,{\mbox{cm}}^3 \, {\mbox{s}}^{-1}$.
It receives contributions from annihilation processes of DM into SM fermions and, provided that the channel is kinematically open,  $aa$ pairs. Expanding $\langle \sigma v \rangle $ in velocity the leading  contribution to the annihilation rate into SM fermions reads:
\begin{align}
\label{eq:sigmavff}
& \langle \sigma v \rangle{(\bar{\chi}\chi \to  \bar{f}f)} \approx \sum_f 
\frac{2 n_c^f c_a^2 m_\chi^2 {g_\chi}^2}{\pi (4 m_\chi^2-m_a^2)^2}\frac{m_f^2}{v_h^2} \nonumber\\
& \approx \left \{
\begin{array}{cc}
7 \times 10^{-23}\,{\mbox{cm}}^3 \, {\mbox{s}}^{-1} g_\chi^2 c_a^2 {\left(\frac{100 \,{\rm GeV}}{m_\chi}\right)}^2 & m_a \ll m_t < m_\chi \\
3\times10^{-26}\,{\mbox{cm}}^3 \,  {\mbox{s}}^{-1} g_\chi^2 c_a^2 {\left(\frac{100\, {\rm GeV}}{m_\chi}\right)}^2 & m_a \ll m_b < m_\chi < m_t
\end{array}
\right.
\end{align}
Here the sum runs over the kinematically accessible SM fermions, $m_{f}$ denotes their mass and $n_c^f$ is their respective color factor. Due to the Yukawa-like coupling structure, the annihilation rate is dominated by the heaviest accessible fermion. In the parameter space of interest here this is either the bottom or the top quark. For convenience we also report two numerical estimates for the cases, $m_b < m_\chi < m_t$ and $m_\chi > m_t$. As can be seen,  DM annihilations into top quarks are extremely efficient so that at least one of the couplings $g_\chi$ and $c_a$ should be substantially smaller than 1 in order to comply with the DM relic density constraint. For $m_\chi < m_t$, in contrast, both $g_\chi$ and $c_a$  need to be order one to achieve a viable thermal DM candidate.

The leading contribution to the other relevant annihilation cross section $\langle \sigma v \rangle (\chi \chi \rightarrow aa)$ only arises at $\mathcal{O}(v^2)$, i.e. it is p-wave suppressed. For $m_a \ll m_\chi$ it can be approximated as 
\begin{equation}
\langle \sigma v \rangle{(\bar{\chi} \chi \to aa)} \approx \frac{g_\chi^4}{192 \pi m_\chi^2}v^2 \approx 4.6 \times 10^{-25}\,{\mbox{cm}}^3 \,{\mbox{s}}^{-1} g_\chi^4 {\left(\frac{100 \, {\rm GeV}}{m_\chi}\right)}^2  
\end{equation}
where we take $v \approx 0.3$ in the second step to give a reasonable estimate for the rate at freeze-out. 
In this regime the DM relic density can in principle be set by the annihilation cross section into $aa$ pairs and the DM relic density  and  $c_a$ could be very small since $\langle \sigma v \rangle{(\bar{\chi} \chi \to aa)}$ is only a  function of $m_\chi$ and $g_\chi$.

In order to relate the DM annihilation cross section with its relic density we have adopted the following relation~\cite{Gondolo:1990dk}:
\begin{equation}
\Omega_\chi h^2=8.76 \times 10^{-11}\,{\mbox{GeV}}^{-2} {\left[\int_{T_f}^{T_0}g_{*}^{1/2} \langle \sigma v \rangle \frac{dT}{m_\chi}\right]}^{-1}
\end{equation}
where $T_f$ and $T_0$ represent, respectively, the standard freeze-out and the present time temperature while $\langle \sigma v \rangle$ is the total cross, i.e.\ summed of all the kinematically accessible final states, DM annihilation cross section. This has been determined by numerically evaluating the integral:
\begin{equation}
\langle \sigma v \rangle=\frac{1}{8 T m_\chi^4 K_2 (m_\chi/T)^2}\int_{4m_\chi^2}^{\infty}ds \sqrt{s}(s-4 m_\chi^2) \sigma (s) K_1\left(\frac{\sqrt{s}}{T}\right)
\end{equation}
where $K_i$ denotes the Bessel function of $i$-th type. The velocity expansion presented above  is only given for illustration, we do not  use it in our computation. 

The freeze-out temperature can be determined by numerically solving:
\begin{equation}
\sqrt{\frac{\pi}{45}}M_{\rm Pl}\frac{45 g_\chi}{4 \pi^4}\frac{K_2 (x)}{h_{\rm eff}(T)}g_{*}^{1/2}m_\chi \langle \sigma v \rangle \delta (\delta+2)=\frac{K_1(x)}{K_2(x)}-\frac{1}{x}\frac{d\log h_{r\rm eff}(T)}{d\log T},\,\,\,\,x=\frac{m_\chi}{T},\delta=1.5
\end{equation}
for $x_f=m_\chi/T_f$. $h_{\rm eff}$ represents the effective entropy degrees of freedom, $g_\chi$  are the internal degrees of freedom of the DM while $M_{\rm Pl}\approx 1.22 \times 10^{19},\mbox{GeV}$ is the Planck mass.
Our results have been double checked with the numerical package micrOMEGAs~\cite{Barducci:2016pcb}.

Today the velocity of dark matter particles bound gravitationally in galaxies is limited by the escape velocity which implies $v=\mathcal{O}(10^{-3})$. Therefore, the rate of dark matter annihilations to pseudoscalar pairs  $\langle \sigma v (\chi \chi \rightarrow a a)\rangle$ is completely negligible at present.
In contrast, the rate for annihilation to $b\bar{b}$ final states is velocity independent and we expect that thermally produced dark matter has $\langle \sigma v \rangle= \mathcal{O}(10^{-26}) \, \mbox{cm}^3 \mbox{s}^{-1}$, see above. Gamma-ray observations in the direction of dwarf Spheroidal galaxies (dSphs) performed by the Fermi-LAT telescope provide strong constraints on  $\langle \sigma v (\chi \chi \rightarrow b \bar{b}) \rangle$  and essentially preclude this annihilation channel from dominating dark matter freeze-out for $m_\chi \lesssim 100$ GeV~\cite{Ackermann:2015zua,Ackermann:2015lka}. A complementary probe will be provided by the Cherenkov Telescope Array which has the potential to exclude the same annihilation cross section for masses above $200$~GeV~\cite{Acharya:2017ttl}.

\subsection{Direct Detection}

\begin{figure}[t]
\begin{center}
\includegraphics[width=0.4\linewidth]{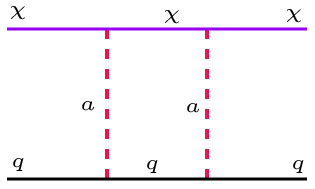}
\end{center}
\caption{Box diagram inducing the SI direct detection cross section in the simplified model.}
\label{fig:feynbox}
\end{figure}

In the scenario under consideration here the prospects for direct detection are generally considered to be rather poor since the effective interaction between a fermion DM candidate and nucleons generated by pseudoscalar exchange is suppressed by the momentum transfer, $\sigma \propto q^4$. Moreover, the cross section is spin-dependent. 
The direct detection phenomenology has been studied in detail for example in~\cite{Arina:2014yna,Dolan:2014ska,Yang:2016wrl}.

The goal of this work is to reconsider direct detection including higher-order effects. In particular, box diagrams with one SM quark and two pseudoscalar states~\cite{Freytsis:2010ne,Ipek:2014gua} running in the loop generate a Spin-Independent (SI) interaction, see Fig.~\ref{fig:feynbox} for a representative diagram.
Despite its origin at higher order, the scattering rate induced by this SI interaction is not necessarily suppressed with respect to the one originating from the tree level pseudoscalar exchange since there is no momentum suppression. On the contrary, an $A^2$ enhancement due to the coherent character of SI interactions increases the experimental sensitivity even further.

The scattering amplitude can be computed starting from the following effective Lagrangian~\cite{Drees:1993bu}:
\begin{equation}
\label{eq:effective_lag_fin}
\mathcal{L}=g_\chi^2 c_a^2 \sum_{q}\frac{m_q^2}{v^2_h}(  C_{V,q}\bar \chi \gamma^\mu \chi \bar{q} \gamma_\mu q+ C_{S,q} \bar \chi \chi \bar{q} q). 
\end{equation}where $q$ is a SM quark and the sum runs over all quark species. 

The coupling of $a$ to the light quarks is highly suppressed due to the SM Yukawa-like coupling structure and, therefore, the only sizable contribution to the direct detection rate is expected from the heavy quarks. This allows for an instant simplification since only valence quarks contribute to the nuclear expectation value of the vector current and we can drop the vector piece in the effective Lagrangian right away. The contribution from the scalar piece is more subtle. Top, bottom and charm quark are clearly heavier than the proton and should be integrated out of an effective theory that describes physics at the nuclear scale. This can be done by invoking the relation between the heavy quark content of the nucleus and the gluon condensate given by~\cite{Shifman:1978zn}
\begin{align}
 m_Q \bar Q Q = -\frac{\alpha_s}{12 \pi}  G_{\mu\nu} G^{\mu \nu},
\label{eq:HeavyQuarkRelation}
\end{align}where $\alpha_s$ is the strong coupling constant and $G_{\mu \nu}$ denotes the field strength tensor of QCD. The numerical value for the gluon condensate is given by $\alpha_s \langle n| G_{\mu \nu} G^{\mu \nu}|n \rangle = -\frac{8}{9} m_N f_{TG}$ with $ f_{TG} \approx 0.894$, see~\cite{Ellis:2008hf} for a detailed discussion of how $f_{TG}$ is extracted from low energy data\footnote{We adopt the default value used by micrOMEGAs~\cite{Barducci:2016pcb}.}.

\begin{figure}
\begin{center}
\includegraphics[width=0.85\linewidth]{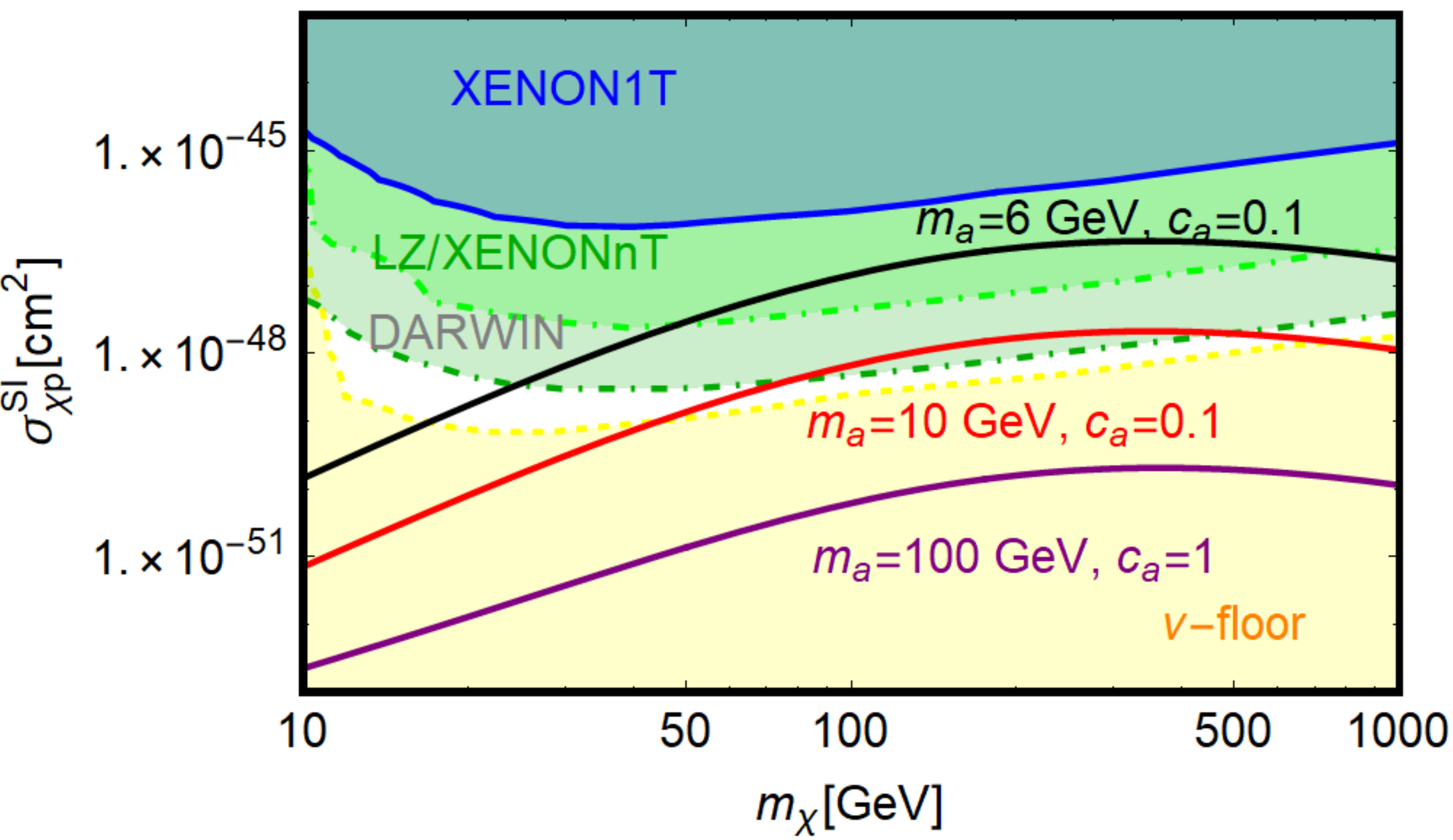}
\end{center}
\caption{SI cross section induced at one-loop as function of the DM mass $m_\chi$, for $g_\chi=0.5$ and for four assignations of $(m_a,c_a)$, as reported on the plot. The blue region is currently excluded by XENON1T. The light green (dark green) region will be probed by LZ and XENONnT (DARWIN). The yellow region corresponds to the sensitivity to coherent scattering processes of neutrinos on nucleons.}
\label{fig:pcross}
\end{figure}

The corresponding SI cross section (for definiteness we will consider the case of scattering on protons) can be schematically expressed as:
\begin{equation}
\sigma_{\chi p}^{\rm SI}=\frac{\mu_\chi^2}{\pi}c_a^4 g_\chi^4 |F_l(m_\chi,m_a)|^2, 
\end{equation}
where $\mu_\chi$ is the reduced mass while:
\begin{equation}
\label{eq:loop_cross}
F_l(m_\chi,m_a)= \frac{2}{27} f_{TG} \sum_{q}\frac{ m_q m_p}{v_h^2} C_{S,q}.
\end{equation}
The expression for $C_{S,q}$ is rather lengthy and we do not report it here but refer the reader to App.~\ref{appx:DDloop} instead.

\begin{figure}
\begin{center}
\includegraphics[width=0.8\linewidth]{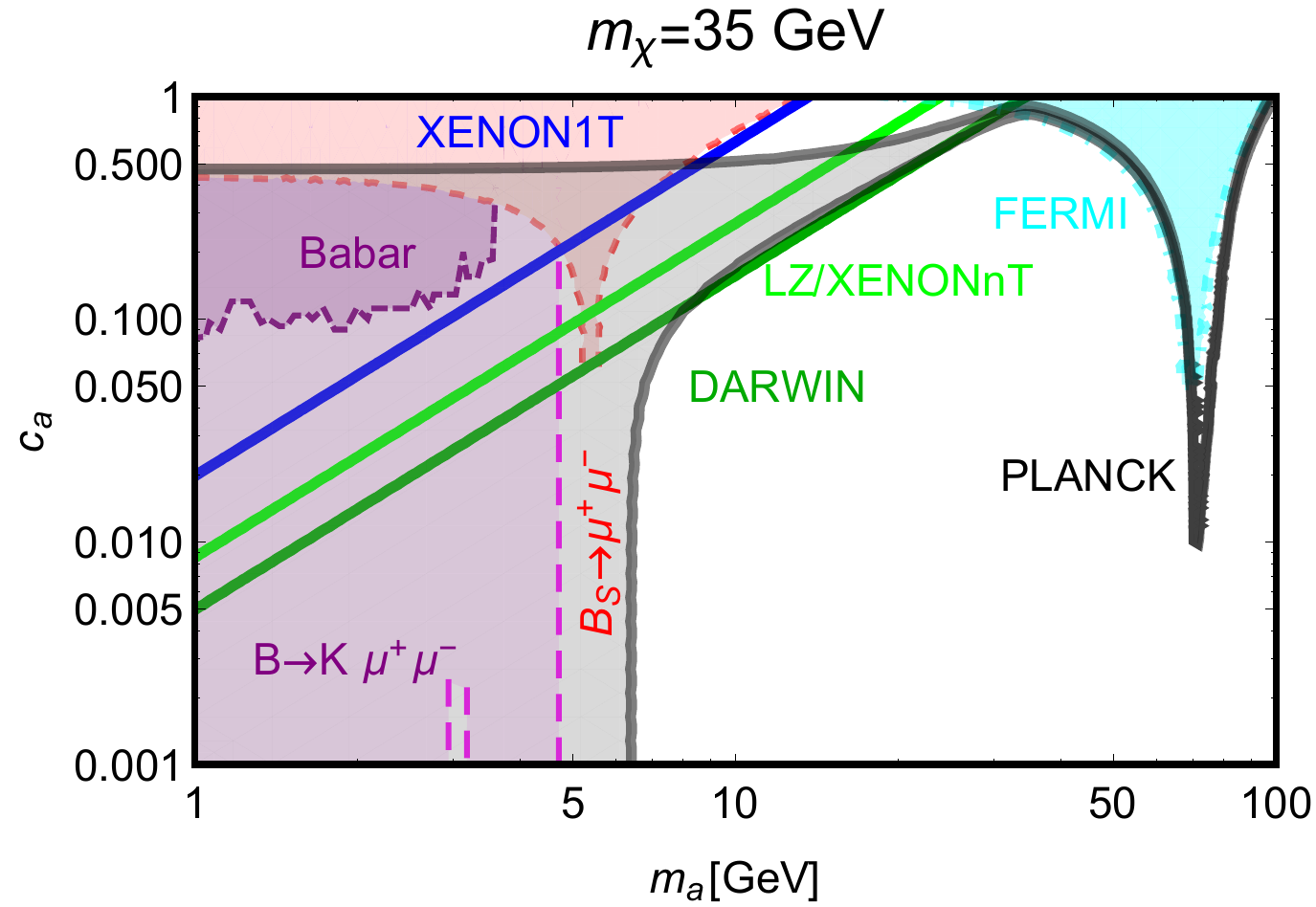}
\end{center}
  \caption{Comparison of various constrains and direct detection  prospects in the $m_a -c_a$ plane for $m_\chi=35$ GeV and $g_\chi= 0.5 $. The direct detection limit from XENON1T is shown in blue while the prospects for LZ and XENONnT (DARWIN) are shown in light (dark) green (Regions above the lines are/will be excluded in absence of detection). Indirect detection limits exclude the cyan region while the relic density constraint can be fulfilled in the gray band.  The bounds from BaBar are depicted in purple  whereas $B
  \rightarrow K \mu^+ \mu^- $ is shown in magenta and $B_s\rightarrow \mu^+ \mu^-$ in red. }
\label{fig:sampleSI}
\end{figure}

Two comments about the reliability of this result are in order. First, the simplified model in which this computation has been made is not gauge invariant. Generically we expect that an UV-completion of the simplified model will introduce new degrees of freedom to restore gauge-invariance. These new fields could allow for additional diagrams and therefore the amplitude considered here cannot be expected to be the full result. We will comment on this in more detail in Sec.~\ref{sec:GaugeInvariant} where we analyze a representative example of such an UV-completion. In addition, there is a further complication which is related to the relation we employed to replace the heavy quarks with the gluons in Eq.~(\ref{eq:HeavyQuarkRelation}). This procedure is justified if the loop that generates the four-fermion interaction and the loop that relates the quarks to the gluon-condensate factorize. While this assumption is reasonable for heavy new physics which can be integrated out at energies above the top mass,  it is not fully appropriate in the scenario under scrutiny here since we are interested in  $m_a < m_t$. In this case, the correct top mass dependence of the effective dark matter gluon interaction is only recovered by a two-loop computation of the effective dark matter gluon interaction~\cite{Hisano:2010ct} which is beyond the scope of this work. In the following, we will rely on Eq.~(\ref{eq:HeavyQuarkRelation}) while keeping in mind that the result is only approximate.

The behavior of the scattering cross section, as a function of the DM mass, for $g_\chi=0.5$ and for some different assignations of $c_a$ and $m_a$, is reported in Fig.~\ref{fig:pcross}. The predictions of the scattering cross sections are compared with the current exclusion limit, as set by XENON1T~\cite{Aprile:2017iyp}, and the projected sensitivities of future  experiments, i.e. XENONnT~\cite{Aprile:2015uzo}, LZ~\cite{Szydagis:2016few} and DARWIN~\cite{Aalbers:2016jon}. Since the expected
sensitivity of LZ and XENONnT are quite similar we only show
one line to improve the readability. The plot also reports the so-called ``neutrino floor''~\cite{Billard:2013qya}, which corresponds to the sensitivity of direct detection experiments to coherent scatterings of neutrinos with nuclei. 

\subsection{Constraints from low energy observables}
\label{sec:SimpleFlavor}
A light pseudoscalar field, $m_a \lesssim 10\,\mbox{GeV}$ can influence a broad variety of low energy observables. For example, it can lead to sizable enhancements of the decay rates of $K$ and $B$ mesons either due to tree-level $a$ exchange or due to the loop induced flavor changing neutral current (FCNC) $b \rightarrow sa$ and $s \rightarrow da$ transitions~\cite{Fayet:2007ua,Andreas:2010ms}. This can change the branching ratios of the aforementioned mesons into lighter mesons and/or leptons.  These limit will be most stringent for light pseudoscalar masses, i.e.\ below the masses of the $B$ and/or $K$ mesons, when $a$ can be produced on-shell in the decay processes.  
An extensive list of constrains on light pseudoscalars has been presented e.g.\  in~\cite{Dolan:2014ska}. Since we are only interested in $m_a > 1 $ GeV we will focus on three  of those processes $\Upsilon\rightarrow a \gamma$, $B_s \rightarrow \mu^+ \mu^-$ and $B\rightarrow K \mu^+ \mu^-$.

The first process is the radiative tree-level decay $\Upsilon \rightarrow \gamma a$ followed by the subsequent decay of $a$ to SM particles. Searches for these decays have been performed by the BaBar collaboration~\cite{Lees:2011wb,Lees:2012iw,Lees:2012te}.  Depending on the mass of $a$   hadronic decays or the leptonic final states $\tau^{+}\tau^{-}$ and $\mu^{+}\mu^{-}$ pose the strongest constraints on the model. In the following we merge these bounds into a single limit which we label BaBar for simplicity.

 The branching fraction for the next process, $B_s \rightarrow \mu^+ \mu^-$, has been measured jointly by the CMS and LHCb collaboration~\cite{CMS:2014xfa} to be $Br\left(B_s \rightarrow \mu^{+} \mu^{-}\right)^{exp}=(2.8^{+0.7}_{-0.6}) \times 10^{-8}$. The experimental value can be related to the theoretical prediction by:
\begin{equation}
Br\left(B_s \rightarrow \mu^{+} \mu^{-}\right)^{exp} \approx \frac{1}{1-y_s} Br\left(B_s \rightarrow \mu^{+} \mu^{-}\right)^{th}
\end{equation}
where the parameter $y_s =\frac{\Delta \Gamma_{B_s}}{2 \Gamma_{B_s}}=0.061$~\cite{Aaij:2014zsa} accounts for the effect of $B_s-\bar B_s$ oscillations. The theoretical prediction of $Br\left(B_s \rightarrow \mu^{+} \mu^{-}\right)$ is given by:
\begin{align}
& Br\left(B_s \rightarrow \mu^{+} \mu^{-}\right)^{th}=\tau_{B_s}\frac{\alpha^2 G_F^2 m_{B_s}}{16 \pi^3}\sqrt{1-\frac{4 m_\mu^2}{m_{B_s}^2}}|V_{tb }V_{ts}|^2 f_{B_s}^2 m_\mu^2  \nonumber\\
& \times |C_{10}^{SM}+\frac{m_{B_s}^2}{2 m_\mu (m_b +m_s)}\frac{m_t^2}{\left(m_{B_s}^2-m_a^2\right)} C_{P,a}|^2, 
\end{align}
where $G_F$ is the Fermi constant, $V_{ij}$ are entries of the CKM matrix. Here $f_{B_s}$ is the $B_s$-meson decay constant  while $\tau_{B_s}$ and $m_{B_s}$ denote the lifetime and mass, respectively. The SM contribution  to the effective operator expansion defined  for example in~\cite{Arnan:2017lxi} (see also~\cite{Altmannshofer:2011gn,Li:2014jsa}) is $C_{10}^{SM} \simeq -4.103$.

In the simplified model determining $C_{P,a}$ poses a problem since a direct computation shows that the amplitude  is divergent. This behavior is not surprising given that the model is not UV-complete. In order to get an estimate for the coefficient $C_{P,a}$ we follow the reasoning proposed in~\cite{Dolan:2014ska} and replace the divergence by a cut-off of the form  $\log(\Lambda^2/m_t^2$), where $\Lambda$ should be interpreted as the scale at which the UV-completion of the simplified model cures the divergence. Under these assumption one expects
\begin{equation}
C_{P,a}\approx\frac{c_a^4}{8\sin^2 \theta_W}\log\left(\frac{\Lambda^2}{m_t^2}\right).
\end{equation}
For illustration we adopt the assignation $\Lambda=1\,\mbox{TeV}$. We would like to highlight that a proper assessment of the limit from this kind of flavor violating processes cannot be achieved within a simplified setup. This issue does not subsist in a gauge UV-complete setup and we will discuss a more robust bound in the next section, where a gauge invariant model will be discussed.

\begin{figure}
\begin{center}
\includegraphics[width=0.85\linewidth]{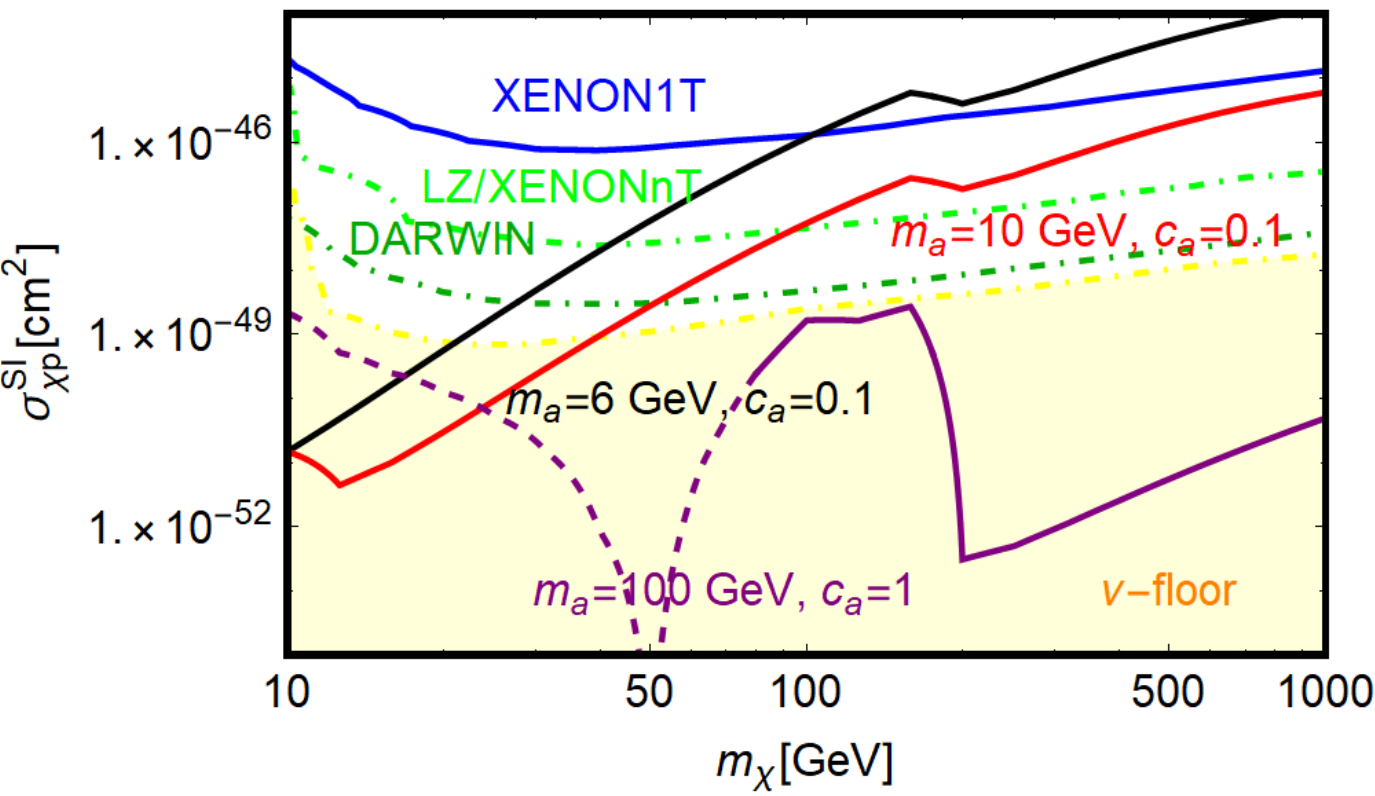}
\end{center}
\caption{Contours of the correct DM relic density, in the  plane $(m_\chi,\sigma_{\chi p}^{\rm SI})$  for three benchmark assignations of $(m_a,c_a)$. The dashed parts of the contours are excluded by indirect detection while the solid parts comply with all constraints. The  dot-dashed light (dark) green line corresponds to the projected sensitivity of LZ and XENONnT (DARWIN). The yellow shaded region is below the so called ``neutrino floor''.}
\label{fig:p1}
\end{figure}

Finally, we  consider the constraint arising from the $B\rightarrow K\mu^{+}\mu^{-}$. Its branching fraction is given by (for simplicity we omit the SM contribution):
\begin{align}
\nonumber
\label{eq:BKmumu_simp}
Br(B \rightarrow K\mu^{+}\mu^{-})=& \tau_B \frac{\alpha^2 G_F^2}{512 \pi^5 m_B^3}|f_0(m_a^2)|^2 m_\mu^2 m_b^2 \sqrt{\lambda(m_B^2,m_K^2,m_a^2)}\\ &\times \frac{m_t^4}{m_W^4}{\left(\frac{m_B^2-m_K^2}{m_b-m_s}\right)}^2 |C_{P,a}|^2 \sqrt{1-\frac{4 m_\mu^2}{m_a^2}}, 
\end{align}
where $\lambda(a,b,c)=(a-b-c)^2-4 b c$ while $f_0$ is a QCD form factor whose most recent numerical determination can be found in~\cite{Bouchard:2013pna,Baru:2015tfa}. 
Similar to the case of $B_s \rightarrow \mu^{+}\mu^{-}$ a well-defined assessment of the experimental constraint requires a UV complete realization but we will again provide an illustrative estimate by taking $\Lambda=1\,\mbox{TeV}$ and compare it with the outcome of the experimental search by LHCb~\cite{Aaij:2012vr}. This comparison is actually less trivial with respect to the case of $B_s \rightarrow \mu^{+}\mu^{-}$. Indeed the observable adopted for experimental analyses is actually $dBr(B \rightarrow K\mu^{+}\mu^{-})/dq^2$ where $q^2$ is the squared invariant mass of the final state muons, rather than the total branching ratio. In order to assess the limit we have imposed that the sum of the SM contribution,  which is obtained by integrating $dBr/dq^2$ (the corresponding expression is found, for example, in~\cite{Arnan:2017lxi}) over the width of the bin, and NP contribution, as given by eq.~\ref{eq:BKmumu_simp} , does not exceed the observed limit in each of the $q^2$ bins of~\cite{Aaij:2012vr} (the NP contribution is more relevant when $a$ decays on shell into a muon pair, i.e. for $q^2=m_a^2$).

\subsection{Results and Discussions}
\label{sec:SimpleDiscussion}

We can now confront direct and indirect detection with low-energy observables and investigate to which extent the loop-induced SI interactions can probe thermal DM. To have an idea of the relative constraining power of the different observables we first show the constraints in the $m_a$, $c_a$ plane in Fig.~\ref{fig:sampleSI}. We have fixed $m_\chi= 35$ GeV and $g_\chi=0.5$ since these parameters do not have an impact on the meson decays. 

As can be seen all limits from meson decays become irrelevant for $m_a \gtrsim 5\,\mbox{GeV}$ with the exception of the one from $B_s \rightarrow \mu^+ \mu^-$. This constraint remains competitive with the  limit from XENON1T  even at higher masses. Next generation experiments provide a better sensitivity; however they can only moderately improve the constraints on $c_a$ because of the very strong dependence, as $c_a^4$, of the scattering cross section. 

The behavior of the relic density band can be understood as follows: For low $m_a$ the full relic density can be explained from $\chi\chi \rightarrow a a $ annihilations alone and $c_a$ has to be smaller than $\sim 0.5$ in order to avoid overproduction. As  $\langle \sigma v (\chi \chi \rightarrow a a)\rangle  $ decreases with increasing $m_a$ an additional contribution from $b \bar{b}$ final states becomes necessary to  produce the correct relic density and $c_a$ is confined to a band. Finally, once annihilations into pseudoscalars cease to be efficient, $b\bar{b}$ remains the only open channel and the allowed range of $c_a$ is constrained to a band that is smaller than the line width in our plot. As expected the annihilation rate features a resonant enhancement when $m_a \approx 2 m_\chi$. 

The Fermi-LAT sensitivity~\footnote{The Fermi-LAT exclusion limit, in Fig.~\ref{fig:sampleSI} and elsewhere in this work has been determined by imposing that the DM annihilation cross section into $\bar b b$ final states, in the $v \rightarrow 0$ limit, is lower than the limit reported in~\cite{Ackermann:2015zua} for this annihilation channel. This procedure is reliable since, given the Yukawa-like couplings of the pseudoscalar mediator the $\bar b b$ is the only relevant SM annihilation final state as long as $m_b < m_\chi < m_t$ (Note however, that hypothetical black hole physics could lead to a dark matter density spike and substantially improving the experimental sensitivity~\cite{Gonzalez-Morales:2014eaa}). 
For $m_\chi > m_t$ DM annihilations are dominated by the $\bar t t$ final state but this is not problematic since indirect detection cannot yet probe the WIMP paradigm at this high DM masses.} is only sufficient to constraint the region where $\chi \chi \rightarrow b\bar{b}$ is the dominant annihilation channel and, therefore, its reach is limited to the resonance region here.  In the future direct detection searches can probe thermal DM for $m_a \lesssim 20\,\mbox{GeV}$ but for $m_a \lesssim 5\,\mbox{GeV}$ the limit from $B_s \rightarrow \mu^+ \mu^-$  forces $c_a$ to be so small that no detection can be expected.

With Fig.~\ref{fig:sampleSI} in mind we can now make an educated guess and select three promising benchmark points (BM1 with $m_a= 6$ GeV, $c_a= 0.1$, BM2 with $m_a= 10$ GeV, $ c_a=0.1$ and BM3 with $m_a= 100$ GeV, $c_a = 1$) for which we analyze the direct detection prospects in more detail. The points are allowed by low energy observables and allow for a successful generation of thermal dark matter. By requiring a thermal dark matter candidate one of the two remaining parameters can be fixed and we can derive the thermal value for $\sigma_{\chi p}^{SI}$ as a function of $m_\chi$, see Fig.~\ref{fig:p1}. 

The shape of the contours can be understood as follows. For the two benchmarks with light $m_a$ and $c_a=0.1$ the DM relic density is mostly determined by the $\bar \chi \chi \rightarrow aa$ process. In this case the correct relic density can be described by a simple relation between $g_\chi$ and $m_\chi$, $g_\chi^2 \approx 0.6 \,
m_\chi/(100\,\mbox{GeV})$~\cite{Dolan:2014ska}. This implies that the predicted scattering cross section increases with the DM mass. The small bumps at $m_\chi \simeq 200 \,\mbox{GeV}$ are induced by a non-negligible contribution  from the $ t \bar{t}$ final state to the DM annihilation cross section. In the case of the benchmark with $c_a=1, \, m_a=100\,\mbox{GeV}$ the relic density is mostly determined by the annihilation into fermion pairs instead. As a consequence we notice two sharp drops in the predicted cross section which are due to the $s$-channel pole, $m_\chi \sim m_a/2$  and to the opening of the $\bar t t$ final state.

As can be seen, the DM scattering cross section of a thermal WIMP is clearly in reach of the next generation of direct detection  facilities for small values of $m_a$, even for $c_a=0.1$ provided  $m_\chi \gtrsim 50\,\mbox{GeV}$. In contrast, for $m_a=100\,\mbox{GeV}$ and $c_a=1$, the predicted cross section lies almost entirely within the ``neutrino floor''. This last benchmark is particularly interesting since it demonstrates the existence of a thermal DM model with a direct detection rate that is  naturally, i.e.\ for $\mathcal{O}(1)$ couplings, below the sensitivity of present and proposed direct detection experiments.\\

\begin{figure}[t]
\begin{center}
\includegraphics[width=0.4\linewidth]{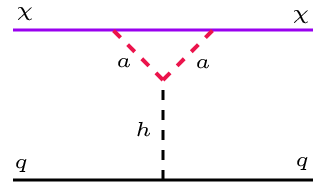}
\end{center}
\caption{Additional diagram contributing of the DM SI cross section if the pseudoscalar mediator couples to the SM Higgs.}
\label{fig:feyntriangle}
\end{figure}

Before moving to a gauge invariant realization of the simplified model we will briefly comment on the changes we expect  if the interaction of the DM or the SM fermions with the mediator are different. As is well known, the case of scalar interactions (i.e.\ without $\gamma_5$) for both DM and SM fermions is strongly constrained since the SI direct detection cross section arises at tree-level. The stringent limits from experiments disfavor values of the mass of the DM and of the mediator below several hundreds of GeV~\cite{Escudero:2016gzx}. 

Alternatively we can consider a Lagrangian  describing a mediator coupled to  a scalar DM current and a pseudoscalar SM fermion current, 
\begin{equation}
\mathcal{L}=g_\chi \bar{\chi} \chi a+i c_a \frac{m_f}{v_h} \bar f \gamma_5 f a\;,
\end{equation}  
or a Lagrangian with a pseudoscalar current for the DM and a scalar current for SM fermions 
\begin{equation}
\mathcal{L}= i g_\chi  \bar{\chi} \gamma_5 \chi a+ c_a \frac{m_f}{v_h} \bar f f a.
\end{equation} 

In the first  case 
the  tree level interaction between the dark matter and nucleons are similar to the pure pseudoscalar case, i.e.\ they are characterized by the same nucleon form factor and there is no coherent enhancement. The scattering cross section is, however, substantially less suppressed and receives an enhancement by a factor $4 m_\chi^2/q^2$, see~\cite{Fitzpatrick:2012ix}. The bounds from low energy observables do not depend on the coupling to the dark matter and therefore they remain unchanged and dominant over direct detection experiments. Contrary to the case studied in the manuscript, the annihilation cross section is $p$-wave dominated and for the same assignation of $(m_\chi,m_a)$  higher values for $g_\chi,c_a$ are required in order to reproduce the relic density. 
Due to the enhancement of the tree-level scattering cross section and the larger expected coupling for freeze-out we find that thermal dark matter with $m_\chi \lesssim 100$ GeV can be ruled out if $m_a\lesssim  10$ GeV and $q \bar{q}$ final states contribute substantially to the relic density. Unfortunately the velocity dependence of the  cross sections reduces the sensitivity of indirect searches and the Fermi-LAT limits from dwarf galaxies are not relevant.

In the second case,
the picture is rather different compared to the case discussed in the paper. In this model the tree-level DM nucleon scattering cross section does not depend on the nucleon  spin and profits from a coherent enhancement. However, the cross section is still suppressed compared to the standard SI interactions since the pseudo-scalar interactions with the DM introduce a factor of $q^2/m_\chi^2$. The DM annihilation cross section into SM fermions is again $s$-wave dominated and, therefore, we expect that the relic density and indirect detection leads to constraints which are similar to the pure pseudoscalar case.
The low-energy bounds which depend on the interactions with the SM fermions have to be reevaluated in this case. Based on results in  the literature, see for example \cite{Schmidt-Hoberg:2013hba,Krnjaic:2015mbs}, we expect the bound to be $c_a < {\mbox{ a few}} \times 10^{-3}$ for $m_a$ below the $B$-meson threshold. This implies  that bounds from low energy observables still dominate over direct detection at low $m_a$.

\section{Gauge Invariant Realization}
\label{sec:GaugeInvariant}

We will now investigate whether the interesting features of the simplified model described above persist in theoretically consistent realizations which respect gauge invariance. One possibility to induce a coupling of the form $a \bar f \gamma_5 f$ between a SM singlet pseudoscalar $a$ and the SM fermions is to mix it with a second pseudoscalar state $A$ which belongs to a Two Higgs Doublet Model (2HDM) extension of the SM~\cite{Ipek:2014gua,Bauer:2017ota,Tunney:2017yfp} (see~\cite{Bell:2016ekl,Bell:2017rgi} for work focusing of mixing between the CP-even scalars). The scalar potential of such a model is given by
\begin{equation}
V=V_{\mbox{\scriptsize  2HDM}}+\frac{1}{2}m_{a_0}a_0^2+\frac{\lambda_a}{4}a_0^4+\left(i \kappa a_0 H^{\dagger}_1H_2+\mbox{h.c.}\right),
\end{equation}
where $V_{\mbox{\scriptsize  2HDM}}$ denotes the usual potential of a 2HDM~\cite{Branco:2011iw}, $H_1$ and $H_2$ are two scalar  SU(2) doublets while  $\kappa$ denotes the coupling between the doublets and the pseudoscalar $a_0$. In the following $\kappa$ is assumed to be real.

Just as in the simplified model the field $a_0$ is coupled to the DM $\chi$ as:
\begin{equation}
\mathcal{L}=i g_\chi a_0 \bar \chi i \gamma^5 \chi\;.
\end{equation}
After EW symmetry breaking the scalar sector of the theory is composed by four CP-even scalars $h,H,H^{\pm}$, and two CP-odd states. The transition from the basis $(H_1,H_2)^{\rm T}$ to $(h,H,H^{\pm},A_0)$ can be expressed in terms of  the angle $\alpha$, which relates the original doublets to the mass eigenstates, and the angle $\beta$, which is given by $\tan\beta=v_2/v_1$ where $v_1$ and $v_2$ are the vacuum expectation values of the two Higgs doublets. A further mixing angle $\theta$ determines the transition from $(A_0,a_0)$ to the basis $(A,a)$ of physical CP-odd eigenstates: 
\begin{equation}
\left(
\begin{array}{c}
A_0 \\
a_0
\end{array}
\right)=
\left(
\begin{array}{cc}
\cos\theta & -\sin\theta \\
\sin\theta & \cos\theta
\end{array}
\right)
\left(
\begin{array}{c}
A \\
a
\end{array}
\right)
\end{equation}
where:
\begin{equation}
\tan2\theta=\frac{2 \kappa v_h}{m_{A_0}^2-m_{a_0}^2}\;.
\end{equation}

Omitting the kinetic terms and the interactions with gauge bosons for simplicity, the interaction Lagrangian in the mass basis reads;
\begin{equation}
\mathcal{L}=\mathcal{L}_{\rm DM}+\mathcal{L}_{\rm scalar}+\mathcal{L}_{\rm Yuk}
\end{equation}
where $\mathcal{L}_{\rm DM}$ is the DM Lagrangian:
\begin{equation}
\mathcal{L}_{\rm DM}=g_\chi \left(\cos\theta a+\sin\theta A\right) \bar \chi i \gamma_5 \chi.
\end{equation}
$\mathcal{L}_{\rm scalar}$ contains the trilinear interactions between the (pseudo)scalar fields
\begin{equation}
\mathcal{L}_{\rm scalar}=\frac{1}{2v_h}\left(m_A^2-m_a^2\right)\left[\sin4\theta aA+\sin^2 2\theta \left(A^2-a^2\right)\right]\left(\sin(\beta-\alpha)h+\cos(\beta-\alpha)H\right),
\end{equation}
while $\mathcal{L}_{\rm Yuk}$ contains the Yukawa interactions with fermions
\begin{equation}
\mathcal{L}_{\rm Yuk}=\sum_f\frac{m_f}{v_h}\left( \xi_f^h h \bar f f+\xi_f^H H \bar f f-i \xi_f^A A \bar f \gamma_5 f-i \xi_f^a a \bar f \gamma_5 a \right).
\end{equation}
The parameters $\xi_f^{\phi}$ with $\phi=h,H,A,a$ depend on the angles $\alpha,\beta$ (as well as $\theta$ for $A,a$) according to the couplings of the original $H_{1,2}$ doublets with the SM fermions. For definiteness, we will consider the type-II 2HDM. Furthermore, we will assume the so-called alignment limit, i.e.\ $\beta-\alpha=\frac{\pi}{2}$, which ensures that the interactions of $h$ are similar to those of the SM-Higgs. Under these assumptions, the scaling factors of the Yukawa couplings are:
\begin{align}
& \xi_f^h=1, \nonumber\\
& \xi_u^H=\frac{1}{\tan\beta},\,\,\,\,\xi_d^H=\xi_e^H=\tan\beta ,\nonumber\\
& \xi_u^A=\frac{\cos\theta}{\tan\beta},\,\,\,\,\xi_d^A=\xi_e^A=\cos\theta \tan\beta ,\nonumber\\
& \xi_u^a=-\frac{\sin\theta}{\tan\beta},\,\,\,\,\xi_d^a=\xi_e^a=-\sin\theta \tan\beta .
\end{align}

As a final simplification we will assume a degenerate spectrum for the scalars and take $m_H=m_A=m_{H^{\pm}}$. The simplified model discussed previously is recovered in the limit $m_\chi, m_a \ll m_A,m_H,m_{H^{\pm}}, \theta \ll 1$ and $\tan\beta=1$. 
However, it should be kept in mind that the heavy Higgs sector cannot be removed completely. 
For a given value of $\theta$, $m_A$ cannot be arbitrarily larger than $m_a$; otherwise violation of unitarity would be encountered in  $aa,aA$ and $AA$ scattering into gauge bosons. The unitarity condition is given by~\cite{Goncalves:2016iyg}:
\begin{align}
\label{eq:uni}
& |\Lambda_{\pm}| \leq 8 \pi \nonumber, \mbox{ where } 
\Lambda_{\pm}=\left[\frac{\Delta_H^2}{v_h^2}-\frac{\Delta^2_a (1-\cos 4\theta)}{8 v_h^2}\pm \sqrt{\frac{\Delta_H^2}{v_h^2}+\frac{\Delta_a^4 (1-\cos4 \theta)}{8 v_h^4}}\right]\nonumber\\
&\mbox{with } \Delta_a^2=m_A^2-m_a^2,\,\,\,\mbox{and }\Delta_H^2=M^2-m_{H^{\pm}}^2+2 m_W^2-m_h^2/2\;.
\end{align}
In the limit $M=m_A=m_{H^{\pm}}\gg m_a$ and taking maximal mixing, i.e.\  $\sin2\theta=1$, this leads to an upper limit on $m_A$ of about $ 1400$ GeV which can be weakened by lowering the value of $\sin\theta$.

The upper limit on the scalar masses should be compared to the lower bounds on the 2HDM from  collider searches~\cite{Khachatryan:2015sma,Aad:2015fna,Aad:2014vgg,Khachatryan:2014wca} and precision observables~\cite{Baak:2011ze,Branco:2011iw}. 
Furthermore, as will be discussed in more detail in the next subsection, the heavy Higgs bosons  contribute substantially to meson decays. In particular the observed branching ratios of weak radiative B-meson decay  impose a  lower bound of approximately 570 GeV~\cite{Misiak:2017bgg} on $m_{H^{\pm}}$ which depends only weakly on $\tan\beta$. 
In addition, searches for the production of $a$ in association with $Z$ and $h$ constrain parts of the parameter space ~\cite{Bauer:2017ota,Tunney:2017yfp}.

Finally, for $m_a \leq m_h/2$, the coupling between the light pseudoscalar $a$ and the SM like-Higgs leads to exotic decays of the SM Higgs bosons. The rate for $h\rightarrow a a $ is given by~\cite{Ipek:2014gua}:
\begin{align}
\Gamma= \frac{(m_A^2-m_a^2)^2}{32 \pi m_h v_h^2}\sin^42\theta \sqrt{1- \frac{4 m_a^2}{m_h^2}}, 
\end{align}
The pseudoscalar can either decay into SM fermions or into a pair of DM states. At the moment the most effective searches have been performed by CMS~\cite{Khachatryan:2017mnf} and rely on the $2b 2\mu$, $4\tau$ and $4\mu$ final states. However, these searches are restricted to specific ranges of $m_a$. In addition, the experimental determination of the Higgs signal strength $\mu \simeq 1.09 \pm 0.11$~\cite{ATLAS-CONF-2015-044} provides an independent constraint of the total width of $h$ into non-standard decay channels.

In the following we will analyze the phenomenology of the gauge invariant model and comment on the differences compared to the simplified model.

\subsection{Dark Matter Annihilations and the Relic Density}

Due to the mixing between the two pseudoscalars $a$ and $A$, the DM annihilations into SM fermions are induced by two mediators. As long as $m_\chi \ll m_A/2$ and $m_\chi < (m_h+m_a)/2$, the DM relic density is controlled by the same processes as in the simplified model. The expression provided in the previous section remain valid provided that the rescaling $c_a \rightarrow \cos\theta \xi^a_f, f=u,d,e$ is used;  similarly the rate for the annihilation into $aa$ final state should be rescaled by a factor $\cos^2\theta$.

When $m_\chi \sim m_A/2$ the DM annihilation cross section is enhanced, with respect to the simplified case, by an additional $s$-channel resonance. Furthermore, as the DM mass increases, new annihilation channels become accessible, namely $ha$, $hZ$, $hA$, $aA$ and $AA$ (the latter two give a suppressed contribution, with respect to the $aa$, since their rates depend on greater powers of $\sin\theta$).

The main difference, concerning the DM relic density between the full and the simplified model, will originate from the $\tan\beta$ dependence of the coupling of the pseudoscalar field with the SM fermions, encoded in the parameters $\xi_f^a$.

\subsection{Direct Detection}

In the case of direct detection, the changes are more striking. As expected, the new fields and interactions introduced in the gauge invariant realization of the pseudoscalar mediator model contribute at the same order as the light pseudoscalar itself and generate additional diagrams. The interaction between $a$ and the Standard Model Higgs $h$ allows for an effective dark-matter-Higgs coupling which is generated by a triangle diagram with pseudoscalars in the loop, see Fig.~\ref{fig:feyntriangle}. This generates a new contribution to SI  cross section. The  triangle contribution to the scalar operator reads~\cite{Ipek:2014gua}:
\begin{equation}
\mathcal{L}=\tilde{C}_S \bar \chi \chi \bar q q,\,\,\,\,\,\tilde{C}_S=\frac{g_\chi^2  \sin ^2 2\theta}{32 \pi^2 m_h^2}\frac{m_A^2-m_a^2}{m_a^2}\frac{m_\chi m_q}{v_h^2} G\left(\frac{m_\chi^2}{m_a^2}\right),
\end{equation}
where the loop function $G(x)$ is given by 
\begin{align}
G(x)=
\frac{ (x-1)\log(x) -2x }{2x^2}+ \frac{(6x-2)\left(\arctan\left(\frac{2x-1}{\sqrt{4x-1}}\right) + \mbox{arccot}\left(\sqrt{4x-1}\right)\right)}{2x^2 \sqrt{4 x-1}}
\end{align}
The new contribution to the scalar coefficient $\tilde{C}_S$ depends on additional parameters of the theory, i.e.\ $m_h$ and $m_A$, and, consequently, the relative importance of the box and the triangle diagram is model dependent. However, the unitarity constraints in the scalar sector ensure that the ratio $m_A/m_h$ can not exceed values of $\mathcal{O}(1)$ unless $\sin{\theta}$ becomes small. In addition, also diagrams with $H$ or $A$ are generated but the larger mass of the heavy scalars suppresses their contribution sufficiently to make them irrelevant for the light $m_a$ scenario under consideration here.

The scalar coefficient for the interaction with the heavy quarks can be related to the coupling with gluons using Eq.~(\ref{eq:HeavyQuarkRelation}) and for the full contribution we sum the box and the triangle induced contributions.\footnote{Since the momentum flow through the Higgs propagator is negligible the two triangle loops can always be factorized and,  in contrast to the box diagram, no subtleties regarding Eq.~(\ref{eq:HeavyQuarkRelation}) arise.  }

\subsection{Constraints from low energy Observables}

As discussed in the previous section, constraints from flavor violating decays of the $B$-mesons cannot be applied easily in a simplified model. In this model, in contrast, it is possible to properly determine the EFT coefficients contributing to their decay rates. The results for  $Br\left(B_s \rightarrow \mu^{+}\mu^{-}\right)$ in the general 2HDM can be written as~\cite{Li:2014jsa,Arnan:2017lxi} 
\begin{align}
 Br\left(B_s \rightarrow \mu^{+} \mu^{-}\right)^{th}  = \;& \tau_{B_s}\frac{\alpha^2 G_F^2 m_{B_s}}{16 \pi^3}\sqrt{1-\frac{4 m_\mu^2}{m_{B_s}^2}}|V_{tb}V_{ts}|^2 f_{B_s}^2 m_\mu^2 \nonumber \\
& \times \left[\big{|} C_{10}+\frac{m_{B_s}^2 C_P}{2 m_\mu (m_b +m_s)}|^2+ |C_S|^2 \frac{m_{B_s}^2 \left(m_{B_s}^2-4 m_\mu^2\right)}{4 m_\mu^2 (m_b+m_s)^2}\right]. 
\end{align}
\noindent
The additional EFT coefficient has to be introduced since all the new scalars contribute non-negligibly to the decay process. For simplicity we report only the effective coefficient associated with the light pseudoscalar field:
\begin{align}
\label{eq:CPa2HDM}
& |C_{P,a}|^2=\frac{m_W^4 \sin^4 \theta}{\left(m_{B_s}^2-m_a^2\right)^2+m_a^2 \Gamma_a^2}|F_P(x_t,x_b,x_\mu,x_{H^{\pm}})|^2,\nonumber
\end{align}
with
\begin{align}
F_P (x_t,x_b,x_\mu,x_{H^{\pm}})= &-\frac{\sqrt{x_b x_\mu} x_t \xi_l}{2 \sin^2 \theta_W}\left \{ \frac{\xi_u^3 x_t}{2}\left[\frac{1}{x_{H^{\pm}}-x_t}-\frac{x_{H}^{\pm}}{(x_{H^{\pm}}-x_t)^2}\log\left(\frac{x_{H^{\pm}}}{x_t}\right)\right]\right.\nonumber\\
&\left. +\frac{\xi_u}{4}\left[-\frac{3 x_{H^{\pm}}x_t-6 x_{H^{\pm}}-2 x_t^2+5 x_t}{(x_t-1)(x_{H^{\pm}}-x_t)}+\frac{x_{H^{\pm}}\left(x_{H^{\pm}}^2-7 x_{H^{\pm}}+6 x_t \right)}{(x_{H^{\pm}}-x_t)^2(x_{H^{\pm}}-1)}\log x_{H^{\pm}}\right.\right.\nonumber\\
&\left. \left. -\frac{x_{H^{\pm}} (x_t^2-2 x_t+4)+3 x_t^2 (2 x_t -2 x_{H^{\pm}}-1)}{(x_{H^{\pm}}-x_t)^2 (x_t-1)^2}\log x_t \right] \right \},
\end{align}
where $x_i =m_i^2/m_W^2,i=\mu, b, t,H^{\pm}$ ~\footnote{Our result disagrees with~\cite{Ipek:2014gua}. This is due to the fact that this reference relies on the computation performed in~\cite{Skiba:1992mg} which found that the branching ratio is enhanced by a factor $\tan\beta^4$. As pointed out by Refs.~\cite{Logan:2000iv,Li:2014jsa}, the result in~\cite{Skiba:1992mg} has been obtained by erroneously omitting relevant diagram and is gauge-dependent. The proper gauge invariant result does not exhibit a $\tan\beta$ enhancement.}.  
For the explicit expressions of the other coefficients generated in the 2HDM we refer  the reader to~\cite{Arnan:2017lxi}. By inspecting Eq.~(\ref{eq:CPa2HDM}) it can be seen that $F_P \propto \frac{1}{2 \sin^2 \theta_W}\log\left(\frac{m_{H^{\pm}}^2}{m_t^2}\right)$ in the limit $m_{H^{\pm}} \rightarrow \infty$, hence recovering the result of the simplified model by identifying $\Lambda=m_{H^{\pm}}$. However, due to the unitarity bound the ratio $m_{H^{\pm}}/m_t$ cannot be arbitrarily large, unless $\sin\theta \rightarrow 0$, so that the SM limit is properly recovered once $m_{H^{\pm}} \rightarrow \infty $. As a consequence one should consider a bound on $B_s \rightarrow \mu^{+}\mu^{-}$ irrespective of the light pseudoscalar. By imposing that the NP contribution maintains the theoretical prediction for $B_s \rightarrow \mu^{+} \mu^{-}$ within $3 \sigma$ with respect to the experimental determination, one excludes values\footnote{This exclusion is approximately the same for all the 2HDM realizations with no flavor changing neutral currents induced a tree level. In the type-II model an additional excluded region appears for $\tan\beta \gtrsim 40$ and $m_{H^{\pm}} \lesssim 200\,\mbox{GeV}$. We will not consider such low values of the mass in our analysis.} of $\tan\beta \lesssim 1$.

The bounds on $Br\left(B \rightarrow K \mu^{+} \mu^{-}\right)$ can been derived analogously. The light pseudoscalar contribution to the effective action is determined by eq.~\ref{eq:CPa2HDM}. This contribution should be complemented by ones depending only on the heavy Higgs states as well as the SM contribution. The full expression of the branching ratio is rather complicated and we will not report it here explicitly and just refer to the literature~\cite{Li:2014jsa,Arnan:2017lxi}.  In order to keep the contributions from the heavy Higgses in agreement with observations, we will impose the bound $\tan\beta \gtrsim 2$ in the following.

The BaBar limits from the $\Upsilon$ decays are far simpler since this is a tree level process. It is sufficient to rescale $c_a^2 \rightarrow \sin^2 \theta \tan^2 \beta \frac{Br\left(a \rightarrow \mu^{+}\mu^{-}(\tau^{+}\tau^{-}\right)_{\rm 2HDM}}{Br\left(a \rightarrow \mu^{+}\mu^{-}(\tau^{+}\tau^{-})\right)_{{\rm simplified}}}$, with $Br()_{\rm 2HDM}$ and $Br()_{\rm simplified}$ being the decay branching ratio of $a$, in the indicated channels, in, respectively, the 2HDM+singlet model and in the simplified model studied in the previous section, and follow the procedure outlined in Sec.~\ref{sec:SimpleFlavor}. 

\begin{figure}[t]
\begin{center}
\subfloat{\includegraphics[width=0.6 \linewidth]{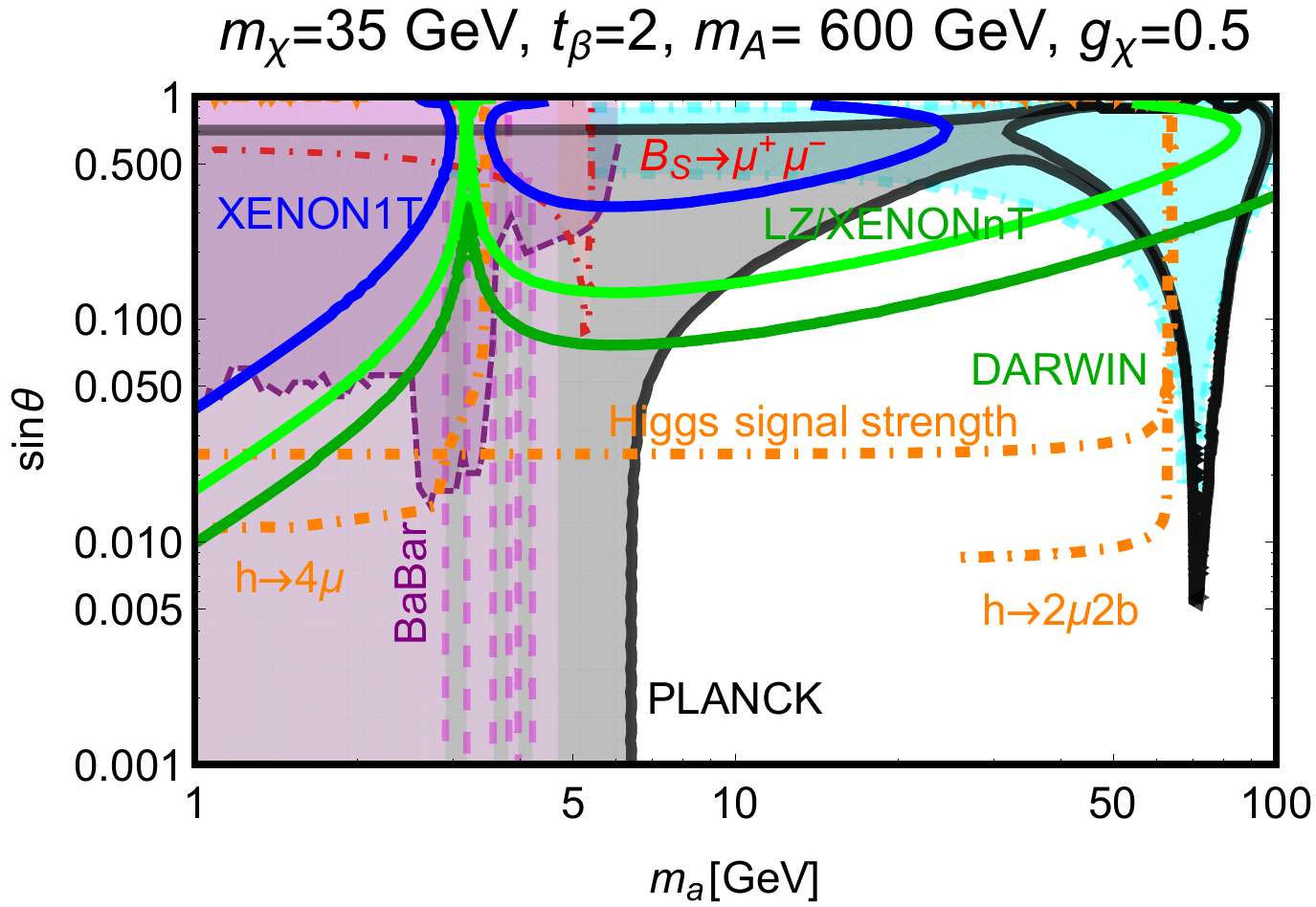}}\\
\subfloat{\includegraphics[width=0.6 \linewidth]{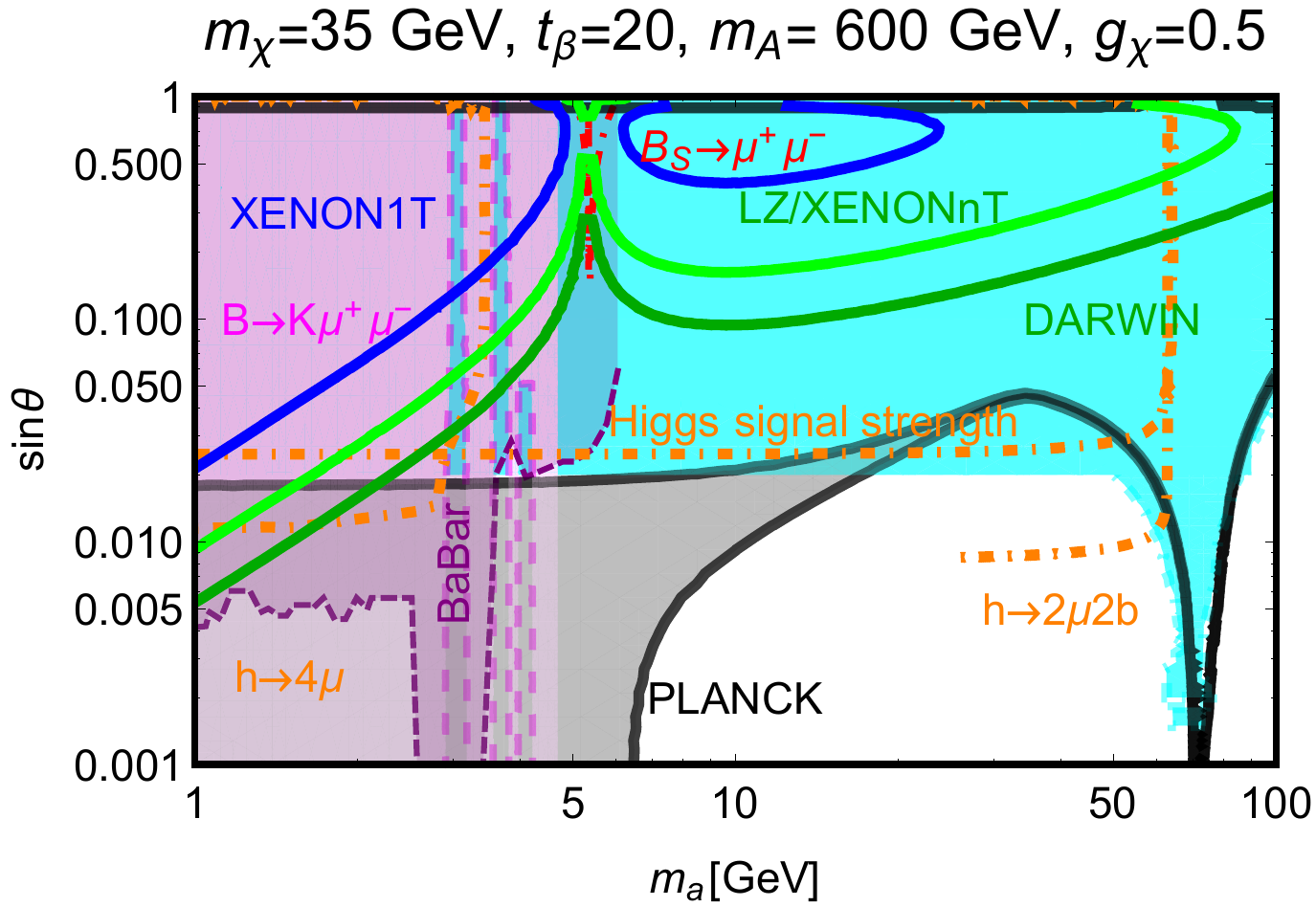}}\\
\subfloat{\includegraphics[width=0.6 \linewidth]{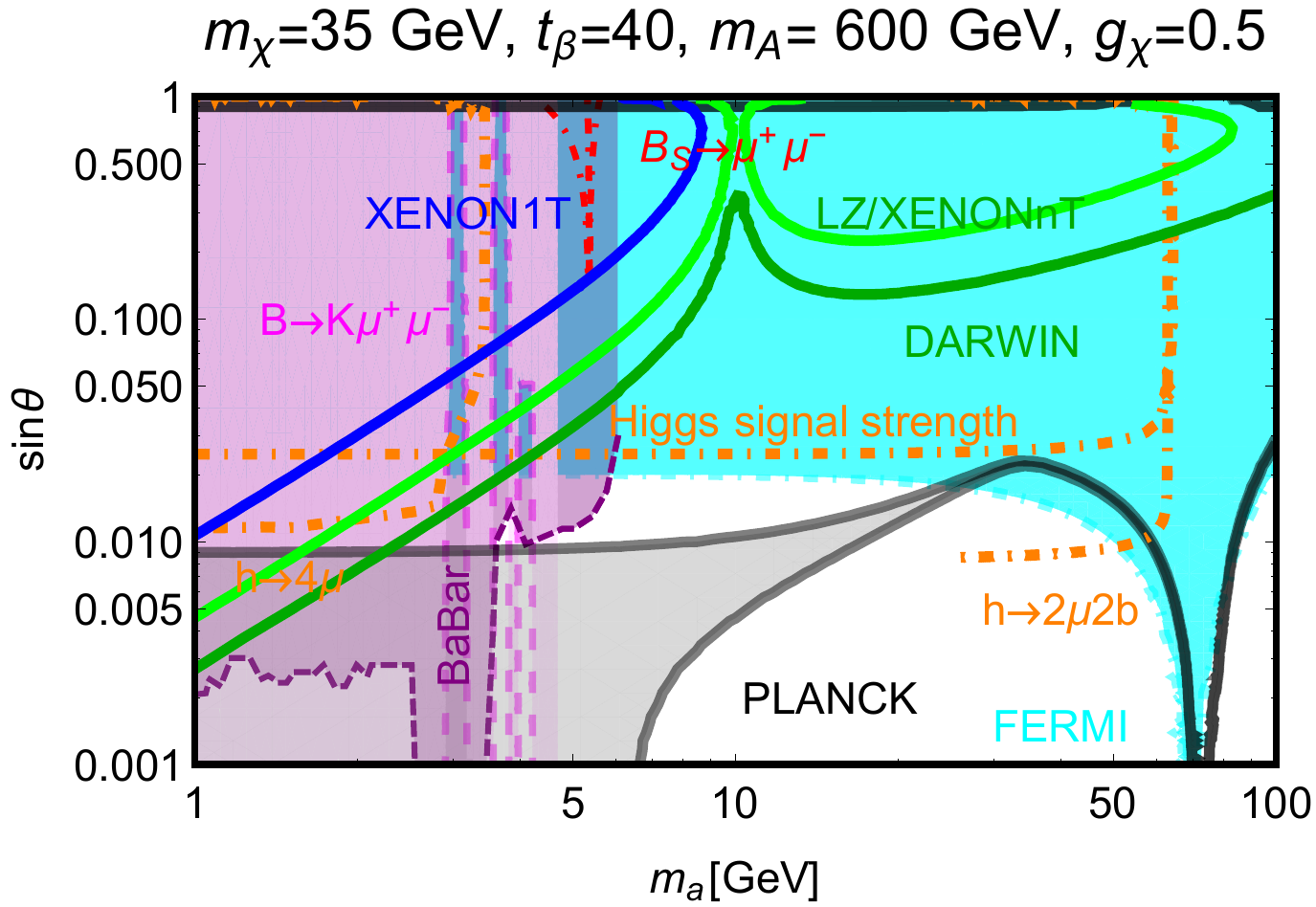}}
\end{center}
\caption{\footnotesize{ Summary of constraints in the plane $(m_a,\theta)$  for fixed assignations of the other parameter, as reported on top of the panels. The correct DM relic density is achieved in the gray region labeled  PLANCK. The regions enclosed in the blue contours are excluded by XENON1T while the ones within the light (dark) green contours correspond to the projected sensitivity of LZ and XENONnT (DARWIN). The cyan region labeled FERMI is excluded by indirect detection. In the red, magenta and purple region, one exceeds the experimental determination of $Br(B_s \rightarrow \mu^+ \mu^-)$ and $Br(B \rightarrow K\mu^+ \mu^-)$  and of the decay rate of $\Upsilon$, respectively. The orange region is excluded by constraints from Higgs decays.} }
\label{fig:pgauge}
\end{figure}

\begin{figure}[t]
\begin{center}
\subfloat{\includegraphics[width=0.6 \linewidth]{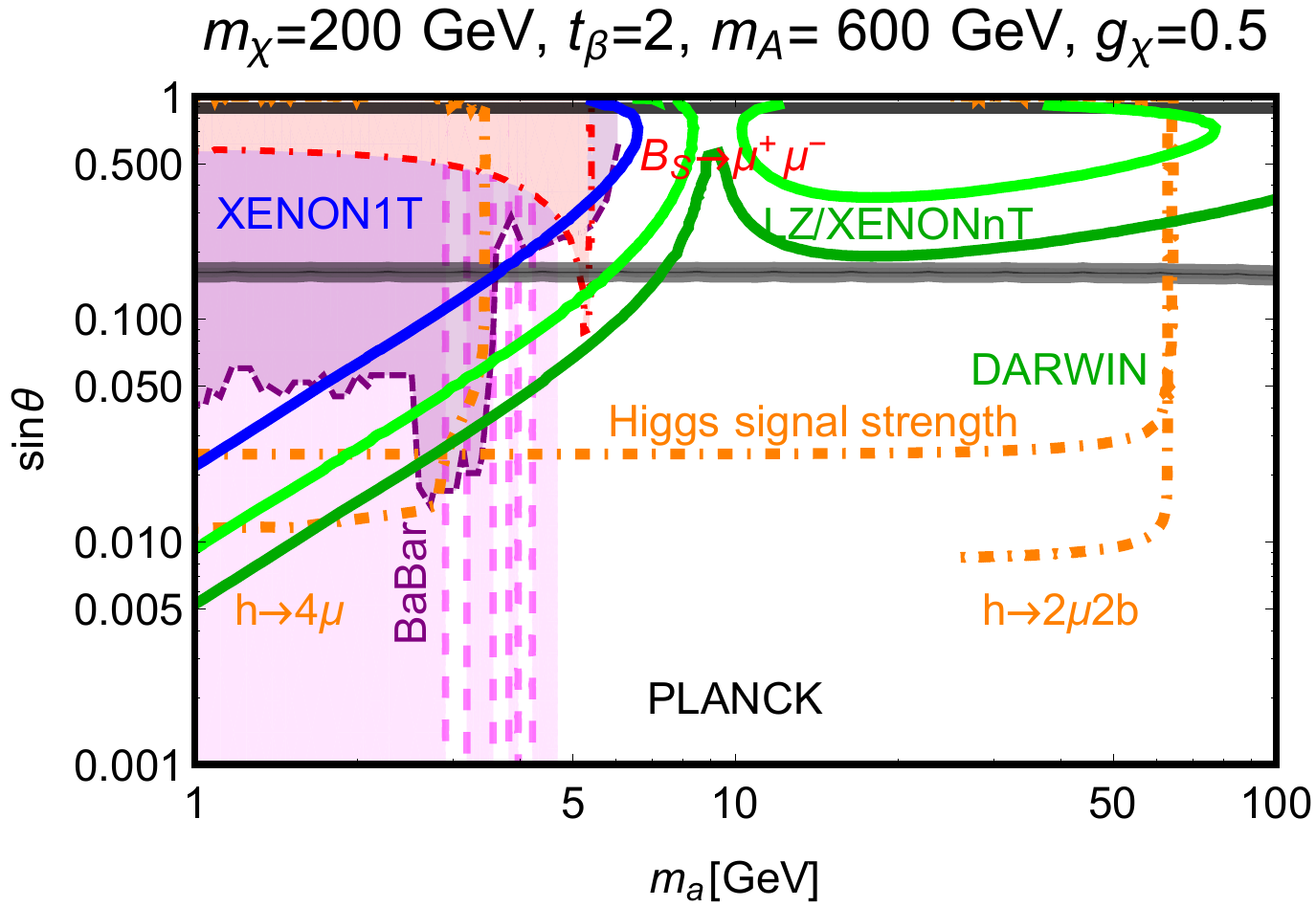}}\\
\subfloat{\includegraphics[width=0.6 \linewidth]{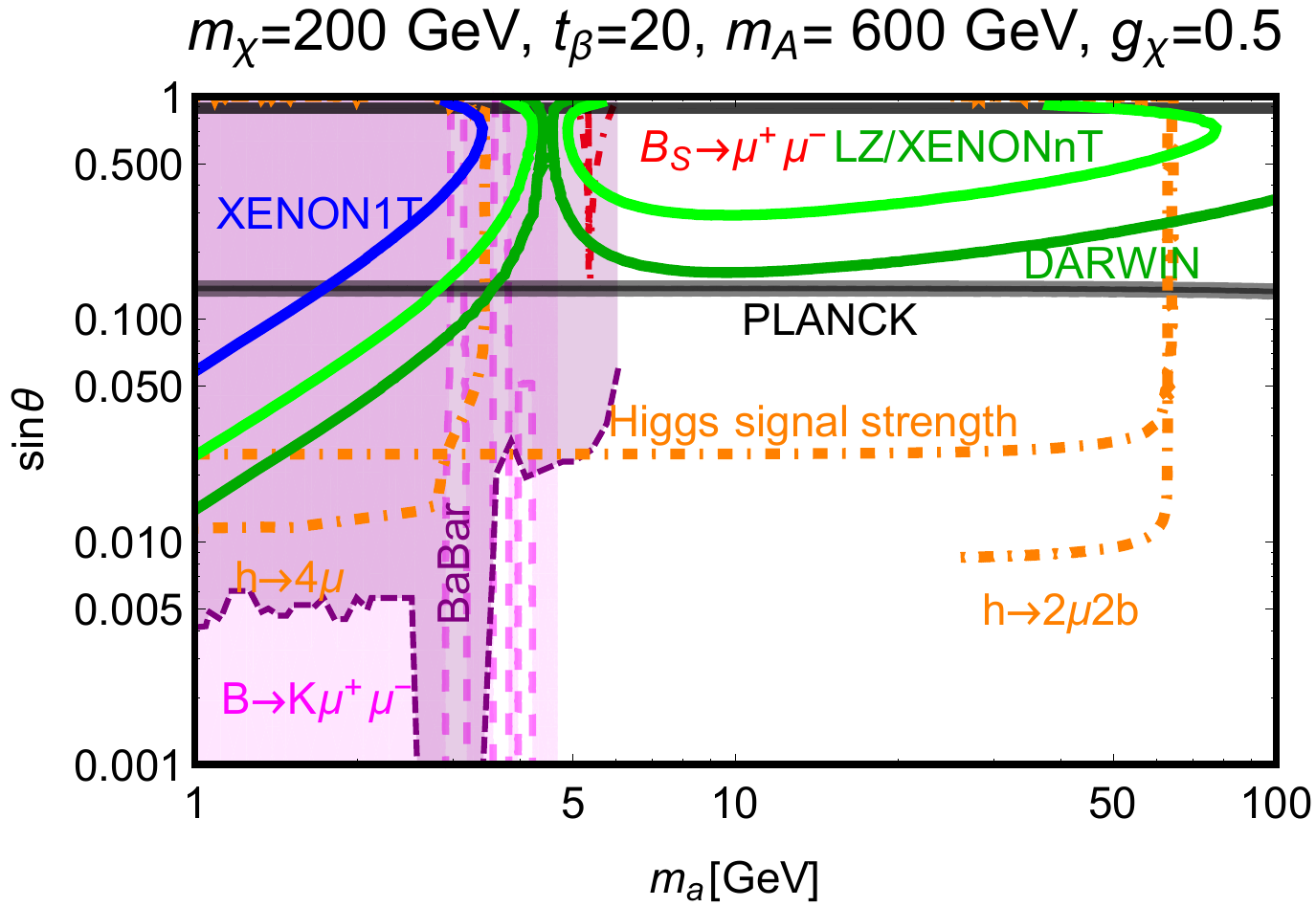}}\\
\subfloat{\includegraphics[width=0.6 \linewidth]{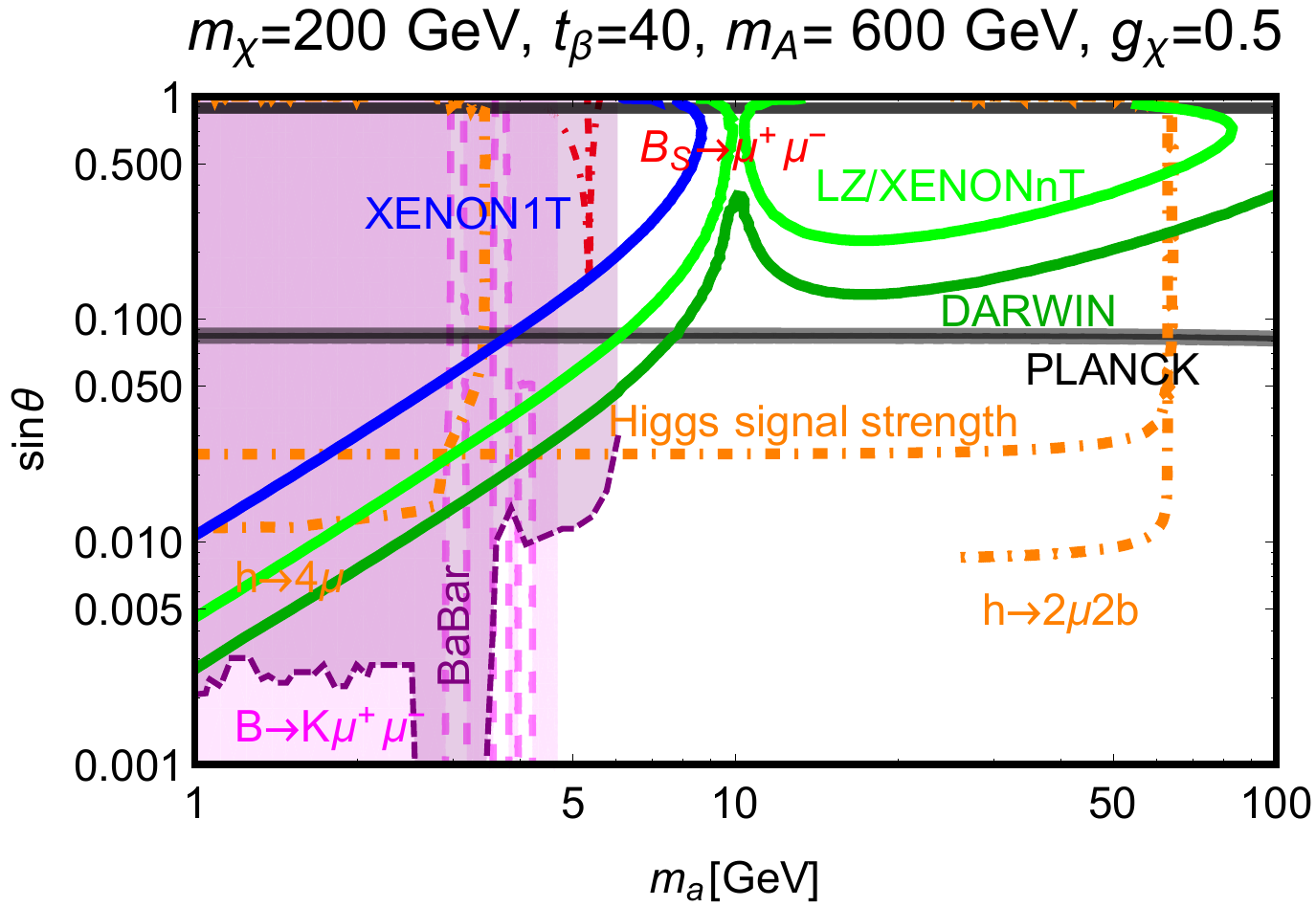}}
\end{center}
\caption{Same as Fig.~\ref{fig:pgauge} for $m_\chi=200\,\mbox{GeV}$.}
\label{fig:pgauge_bis}
\end{figure}

\subsection{Results and Discussion}
Similar to  our discussion of the simplified model we  compare the relative strength of the different constraints in the $m_a$-$\sin \theta$ plane  for $m_\chi=35$  GeV and $g_\chi= 0.5$. We need to keep in mind, however, that the full model has additional parameters which can have an impact on the phenomenology. We set $m_A=m_H=m_{H^\pm}=600 \,\mbox{GeV}$, close to the lower limit imposed by flavor observation and consider three different cases with $\tan\beta =$ 2, 20 and 40. 

Again the gray regions highlight the part of the parameter space that produces the correct relic density. For the DM mass considered here, the main annihilation channels are still into $aa$ and $\bar b b$ finals states. The former mostly dominates at low values of $m_a$. Since for $\sin\theta \ll 1$ the corresponding rate is basically independent of the $\theta$ angle, the correct DM relic density is achieved in a broad region rather than along a narrow contour. At higher values of $m_a$, and in particular for $m_a > m_\chi$ the correct relic density is instead restricted to a narrow band and exhibits the expected pole  at $m_\chi \sim \frac{m_a}{2}$ where the correct relic density is achieved for very low values of $\theta$. Due to the $\tan\beta$ enhancement of the coupling of the pseudoscalar mediator with the b-quark the region of the correct relic density shifts towards lower values of $\theta$ as $\tan\beta$ increases. 

Given  that the DM annihilation cross section into $\bar b b$ final states is s-wave dominated this is accompanied by stronger limits from Fermi-LAT, see the cyan region in Fig.~\ref{fig:pgauge}. Concerning direct detection, the limit/projected sensitivities only show a modest dependence on $\tan\beta$ but the shape of the curves exhibits some clear differences compared to the simplified scenario. This is due to the interplay between the two different contributions to the scalar effective operator. At low $m_a$ the DM scattering cross section is dominated by the box-diagram while the triangle loop gives the largest contribution in the high $m_a$ regime. At intermediate masses, a ``blind spot'' appears since the two contributions interfere destructively. At high values of $\tan\beta$, the region of the viable relic density moves increasingly away from the sensitivity of direct detection facilities. For $\tan\beta >20$ only a small region, corresponding to light $m_a$ masses, can be probed by DARWIN. This region is however completely ruled out by the constraints from  $B \rightarrow K \mu^{+} \mu^{-}$.
A further effective constraint is due to the $h \rightarrow aa$ decay. By considering only the bound from the Higgs signal strength, values of $\sin\theta$ greater than 0.05 are excluded within the full kinematical range of the $h \rightarrow aa$ decay, stronger bounds are obtained for more limited ranges of $m_a$ when one considers searches of specific final states. 
Once this bound is enforced the correct DM relic density for $m_\chi = 35$ GeV can only be achieved in regions of parameter space that are out of reach of direct detection experiments.
In summary, the case of $m_\chi=35\,\mbox{GeV}$ appears to be very strongly constrained. Despite our revision, direct detection prospects remain irrelevant compared to bounds coming from low-energy/collider searches.

We consider a second benchmark with higher DM mass, namely $m_\chi=200\,\mbox{GeV}$. Despite the slightly lower sensitivity from direct detection experiments the high value of the DM mass has two advantages: the DM annihilation cross section is enhanced by the $\bar t t$, $ha$ and $Zh$ channels, so that the correct DM relic density can be achieved for higher values of $m_a$, to which flavor and collider bounds are not sensitive. Bounds from indirect detection are evaded since they cannot probe thermal DM with this such a high the mass yet. The results of our analysis are shown in Fig.~\ref{fig:pgauge_bis}. Contrary to the previous benchmark the relic density is constrained to two narrow lines. The line at $\sin\theta\simeq 1$ corresponds to a relic density mostly determined by annihilations into $\bar t t$ final states while at $\sin\theta \sim 0.1$ the  dominant contribution is due to the $ha$ final state. For $m_a > m_h/2$ the upper curve is unstrained by Higgs decay and can be tested by upcoming direct detection experiments. 

\begin{figure}[t]
\begin{center}
\subfloat{\includegraphics[width=0.6\linewidth]{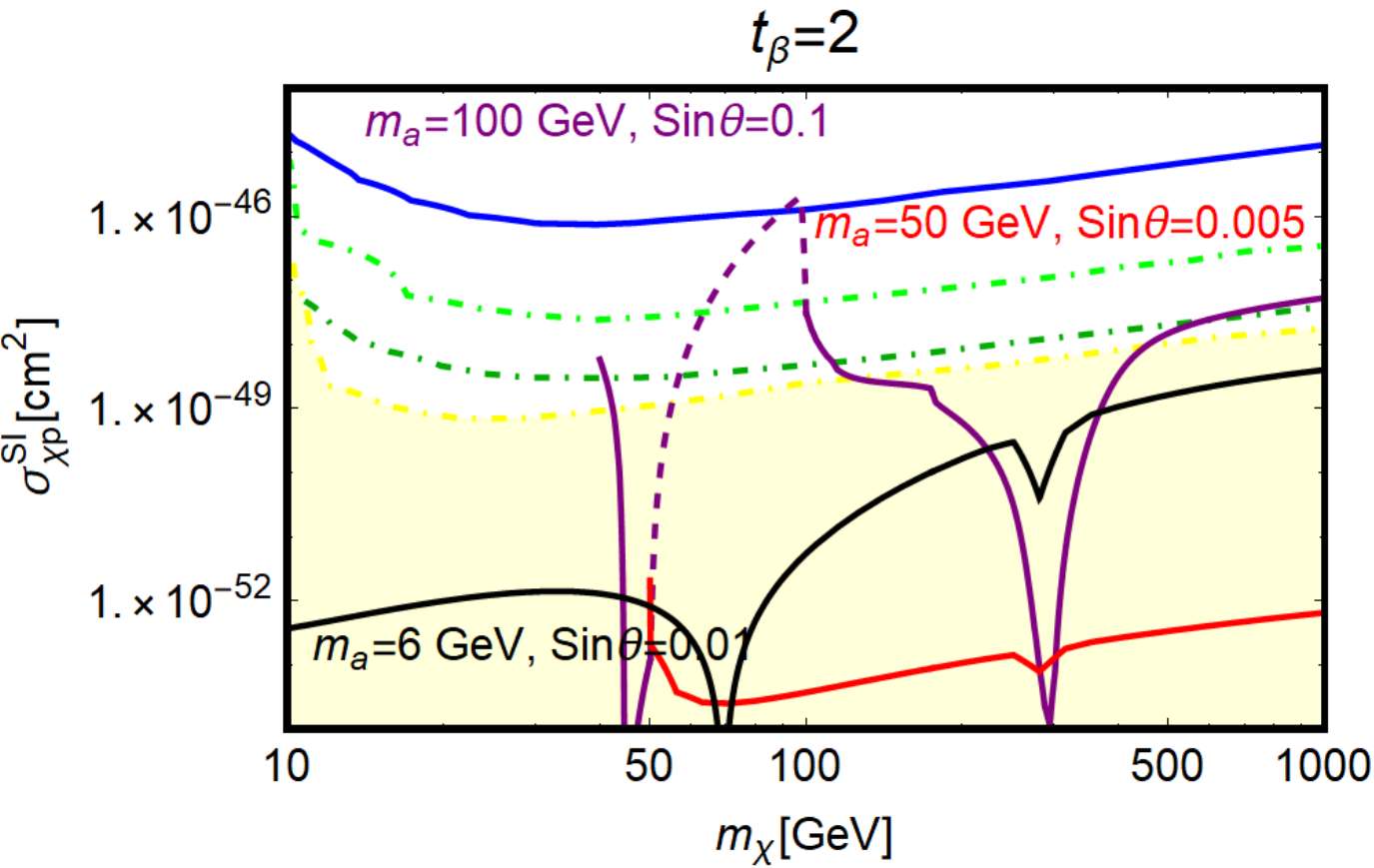}}\\
\subfloat{\includegraphics[width=0.6\linewidth]{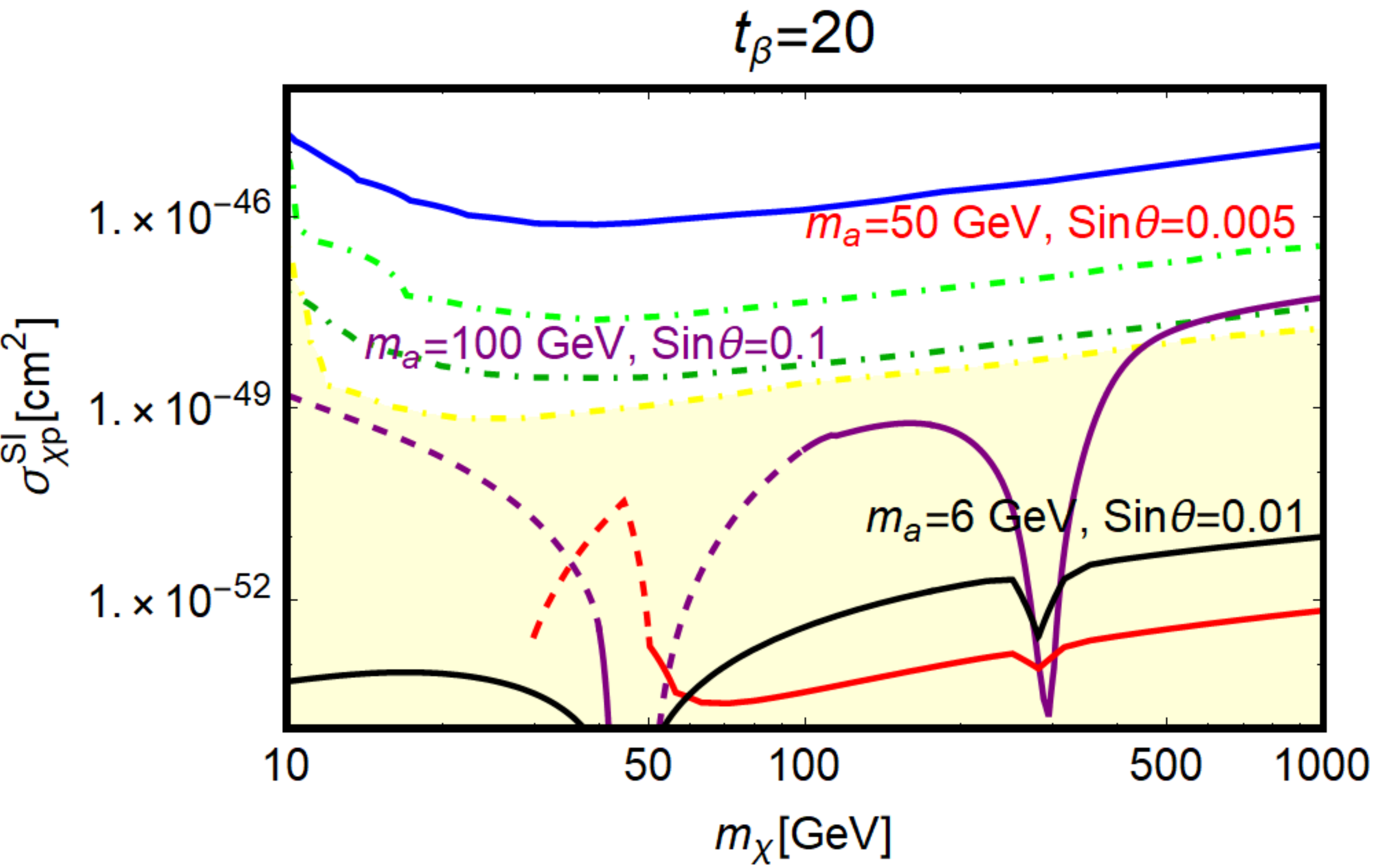}}\\
\subfloat{\includegraphics[width=0.6\linewidth]{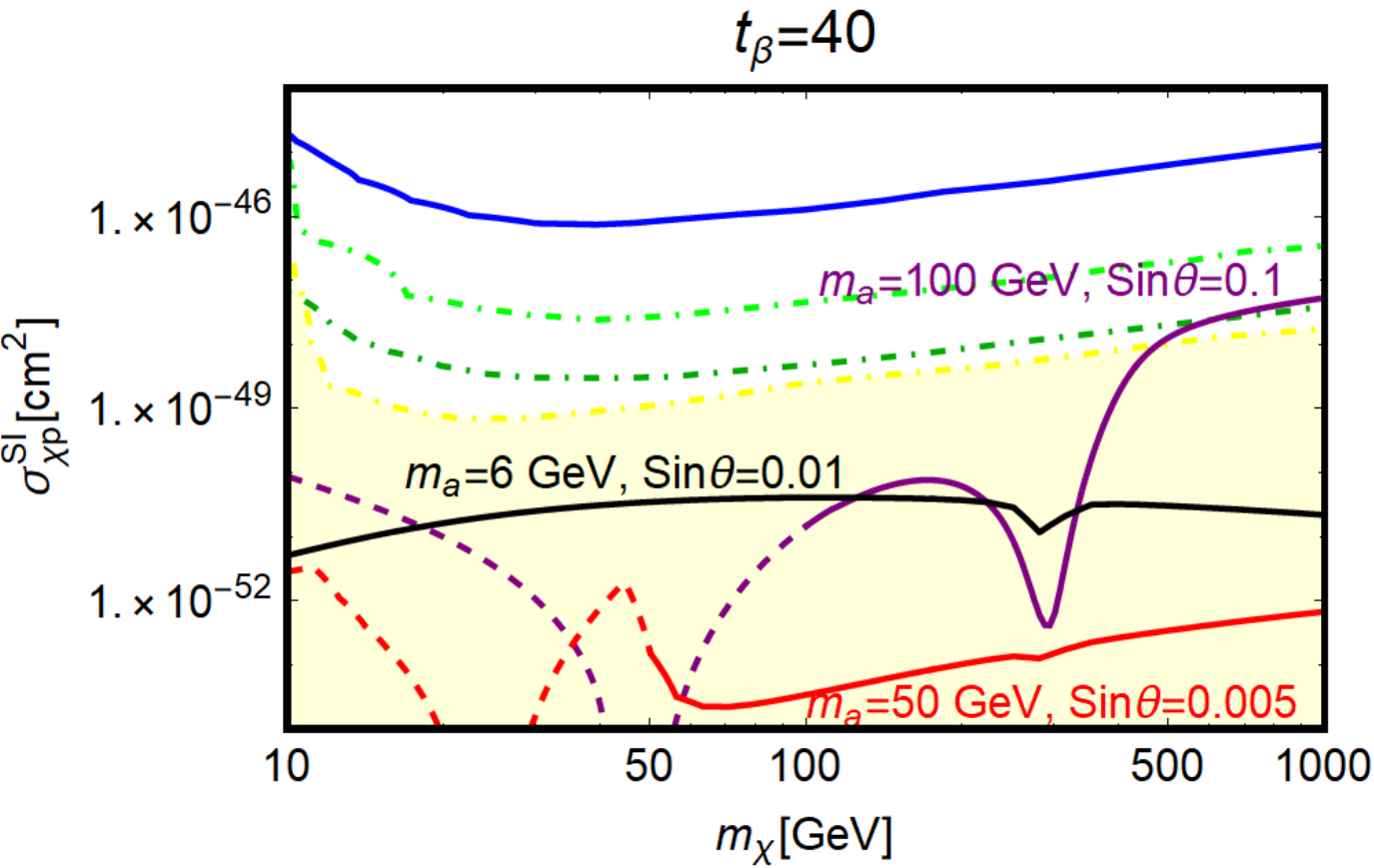}}
\end{center}
\caption{\footnotesize{Predictions for the direct detection cross section of thermal dark matter for three benchmark models with low $\tan \beta$, i.e. $t_\beta=2$ (upper panel), medium $\tan \beta$, i.e. $t_\beta=20$ (central panel), and high $\tan \beta$, i.e. $t_\beta=40$ (lower panel). The solid parts of the lines are allowed by all other constraints while the dashed parts are excluded by indirect detection. In regions where the relic density can not be achieved with perturbative couplings the lines are discontinued.  Also shown are the XENON1T limit (blue, solid), and the projected sensitivity of LZ and XENONnT (light green, dot-dashed) and DARWIN (dark green, dot-dashed). The yellow region is below the neutrino floor.} }
\label{fig:psgima}
\end{figure}

Following the same procedure as in Sec.~\ref{sec:SimpleDiscussion} we select three benchmarks points which are unconstrained by meson and  Higgs decays, BM1 with ($m_a= 6$ GeV and $\sin \theta = 0.01$), BM2 with ($m_a=50$ GeV and $\sin \theta =0.005$) and BM3 with ($m_a= 100$ GeV and $\sin \theta = 0.1$). Each benchmark is considered for $\tan \beta =2$, 20 and 40 while the mass of the heavy Higgses remains fixed at $600$ GeV. We employ the relic density constraint to fix the value of $g_\chi$ as a function of $m_\chi $ and show the predicted values for $\sigma^{SI}_{\chi p}$ in Fig.~\ref{fig:psgima}. In contrast to the simplified model the expected scattering cross section of  thermal dark matter is now a more complicated function of $m_\chi$ and the interplay of the box and the triangle contribution with the relic density constraint introduces a number of dips in $\sigma_{\chi p}^{SI}$. Nevertheless, it is clear that the scattering rate for light mediators is rather suppressed due to the strong bound on $\sin \theta$ from Higgs decays. \FloatBarrier For  $m_a=100$ GeV , on the other hand, detectable values $\sigma_{\chi p}^{\rm SI}$ are still possible  if either $\tan \beta $ is not too large or the annihilation rate is dominated by $t\bar{t}$ final states. \\

Before coming to our conclusions it is worthwhile to revisit the question regarding alternative coupling structures which we briefly discussed at the end of Sec.~\ref{sec:Simple}.   
 Once we introduce such a scalar-pseudoscalar interaction to our model, $CP$ is no longer a good quantum number of our theory. Now all four scalar mass eigenstates can mix and should be considered simultaneously. The mixing pattern depends on the details of the scalar potential and  can not be related to the interactions of the lightest scalar with the DM and the SM fermions in a simple way. A realistic analysis of this model would require that all four scalar mass eigenstates are included both at tree and loop-level. Such an analysis is beyond the scope of our work. Given that the additional interactions of the UV-complete pseudoscalar model changed the picture substantially compared to the simplified model, we expect a similar effect for a model with $CP$ violation and would like to stress, again, that the estimates for the simplified model should be taken with a grain of salt.
 
\section{Conclusions}

In this work, we have re-analyzed the  direct detection prospects for dark matter models with pseudoscalar mediators. Since the tree-level dark matter-nucleon cross section is momentum suppressed the leading contribution to the direct detection rate arises at  higher order. We have calculated the loop-induced contribution to the scattering rate in a  simplified and in a more realistic, gauge-invariant model.

Using the relic density as our guiding principle and taking additional constraints from indirect detection and low energy observables  into account we have identified the most promising regions for future direct detection experiments. 
In light of the constraints from meson decay, the detection of dark matter interacting via a very light pseudoscalar ($m_a\lesssim 5$ GeV) is challenging and can not be expected in the upcoming direct detection experiments. However, in the simplified model, we find promising regions of parameter space with $m_a=\mathcal{O}(10)$ GeV in which thermally produced dark matter can be tested by experiments with the sensitivity of the projected LZ, XENONnT and Darwin detectors. 
For heavier mediators, i.e.\ $m_a \sim 100$ GeV, an observation of dark matter-nucleon scattering might still be possible if the sensitivity can be pushed beyond the neutrino floor. At even higher masses collider searches can be expected to be the most restrictive experiments, again stressing the complementarity between direct detection and the LHC.

In order to assess whether these conclusions are robust, we repeat our analysis in a realistic completion of the simplified models. In the setup considered here, gauge invariance is restored by mixing the mediator with the pseudoscalar component of a 2HDM. Due to the larger scalar sector,  additional interactions arise which change the global picture considerably. In particular, the new coupling between the pseudoscalar and the SM Higgs has an important impact on the phenomenology. On one hand, this coupling leads to decays of the SM-like Higgs $h$ into pairs of pseudoscalars. The rate of such exotic Higgs decays is already tightly constrained by the observed Higgs signal strength and dedicated searches. In light of these limits, a detectable signal arising from the most promising parameter space identified in the simplified model, i.e.\ $m_a \approx 10$ GeV, is essentially ruled out. On the other hand, the same interaction also leads to an additional contribution to the direct detection rate which enhances the DD signal for heavier $m_a$ relative to the simplified model. In particular, at high $m_\chi$ the predicted scattering rate is within reach of a Darwin-like device even for $m_a = 100$ GeV.

To conclude, we want to emphasize that despite their essentially vanishing dark matter-nucleon cross section at tree level, models with pseudoscalar mediators are potentially detectable with the next generation of direct detection experiments. In addition, they achieve such small scattering rates naturally, i.e.\ for $\mathcal{O}(1)$ couplings, and could, therefore, provide an important benchmark for upcoming direct dark matter searches. Finally, we would like to encourage the community to fully exploit the potential of direct detection experiments and think about new ways to extend the sensitivity beyond the neutrino floor. 

\vspace{0.5cm}

\paragraph*{Acknowledgments.---}
The authors thank L.\ Di Luzio and F.\ Mescia for the fruitful discussions. The authors also warmly thank D.\ McKeen and M.\ Freytsis for the kind exchange of correspondence and N. Bell and G. Busoni for helpful comments. FSQ acknowledges support from MEC and ICTP-SAIFR FAPESP grant 2016/01343-7. 
WR is supported by the DFG with grant RO 2516/6-1 in the Heisenberg program.

\appendix

\section{SI Loop computation} 
\label{appx:DDloop}

In this appendix we sketch the derivation of the function $F_l$ which determines the SI cross section, see Eq.~(\ref{eq:loop_cross}). As already mentioned this scattering cross section is associated to the box diagram described in the Fig.~\ref{fig:feynbox}. We have computed the diagram(s) with the package FeynCalc~\cite{Mertig:1990an,Shtabovenko:2016sxi}, upon implementing the simplified model into FeynArts~\cite{Hahn:2000kx}. This computation allows to determine the following effective Lagrangian:
\begin{align}
\label{eq:effective_lagrangian}
& \mathcal{L}=A \bar{u}(p_q^{'})\gamma^\mu u(p_q) \bar{u}(p_\chi^{'})\gamma_\mu u(p_\chi)+B \bar{u}(p_q^{'})u(p_q) \bar{u}(p_\chi^{'})u(p_\chi)\nonumber\\
& +\bar{u}(p_q^{'})\left(C_1 \slashed{p}_\chi+C_2 \slashed{p}_\chi^{'}\right)u(p_q)\bar{u}(p_\chi^{'})u(p_\chi)+D\bar{u}(p_q^{'})\slashed{q}u(p_q)\bar{u}(p_\chi^{'})u(p_\chi)\nonumber\\
& +E \bar{u}(p_q^{'})u(p_q)\bar{u}(p_\chi^{'})\slashed{p}_q^{'}u(p_q)+ \bar{u}(p_q^{'})\left(F_1\slashed{p}_\chi+F_2\slashed{p}_\chi^{'}
\right)u(p_q) \bar{u}(p_\chi^{'}) \slashed{p}_q^{'}u(p_\chi)\nonumber\\
&+G \bar{u}(p_q^{'})\slashed{q}u(p_q) \bar{u}(p_\chi^{'}) \slashed{p}_q^{'}u(p_\chi)
\end{align}
The operators containing $\slashed{q}=\slashed{p}_\chi-\slashed{p}_\chi^{'}=\slashed{p}_q^{'}-\slashed{p}_q$ become null once the Dirac equation is applied. 
By making repeated use of the equations of motion it is possible to reduce the effective Lagrangian~(\ref{eq:effective_lagrangian}) to the sum of only one vectorial and one scalar operator, as written in~(\ref{eq:effective_lag_fin}), whose corresponding coefficients are given by (notice that $C_1$ and $C_2$ cancel each other one the external particles are put on shell): 
\begin{align}
& C_{V,q}=A+m_q E+2 m_\chi m_q (F_1+F_2) \nonumber\\
& C_{S,q}=B
\end{align}
Due to the Yukawa-like coupling structure vector interactions do not contribute to the scattering rate and we do not report them. The coefficient of the scalar coefficient can be evaluated analytically\footnote{We use Package-X~\cite{Patel:2015tea} for the reduction of the Passarino-Veltman functions.} and is given by 
\footnotesize{
\begin{align}
& C_{S,q}=-\frac{1}{960 \pi ^2 m_q^3 m_\chi^3 m_a^4 (m_q-m_\chi)^3
   (m_q+m_\chi)^3}\left[(m_q-m_\chi)^3 \left(m_q^4 m_a^4 \left(5 m_\chi^2 (m_q+3 m_\chi)-m_a^2 (3 m_q+5 m_\chi)\right) \log
   \left(\frac{m_\chi^2}{m_a^2}\right)\right.\right.\nonumber\\
& \left. \left.   -2 m_q^4 \sqrt{m_a^4-4 m_\chi^2 m_a^2} \left(8 m_q m_\chi^4+m_\chi^2 m_a^2 (m_q-5
   m_\chi)+m_a^4 (3 m_q+5 m_\chi)\right) \log \left(\frac{\sqrt{m_a^4-4 m_\chi^2 m_a^2}+m_a^2}{2 m_\chi m_a}\right)\right.\right.\nonumber\\
&\left.\left.+m_\chi^4 m_a^4
   \left(5 m_q^2 (3 m_q+m_\chi)-m_a^2 (5 m_q+3 m_\chi)\right) \log \left(\frac{m_q^2}{m_a^2}\right)\right.\right.\nonumber\\
   &\left.\left.+2 m_q^2 m_\chi^2 m_a^2
   (m_q+m_\chi) \left(8 m_q^2 m_\chi^2-m_a^2 \left(3 m_q^2+2 m_q m_\chi+3 m_\chi^2\right)\right)\right.\right. \nonumber\\
   &\left.\left. -2 m_\chi^4 \sqrt{m_a^4-4 m_q^2
   m_a^2} \left(8 m_q^4 m_\chi+m_q^2 m_a^2 (m_\chi-5 m_q)+m_a^4 (5 m_q+3 m_\chi)\right) \log \left(\frac{\sqrt{m_a^4-4 m_q^2
   m_a^2}+m_a^2}{2 m_q m_a}\right)\right)\right. \nonumber\\
   &\left. +(m_q+m_\chi)^3 \left(m_q^4 m_a^4 \left(5 m_\chi^2 (m_q-3 m_\chi)+m_a^2 (5 m_\chi-3
   m_q)\right) \log \left(\frac{m_\chi^2}{m_a^2}\right)\right.\right. \nonumber\\
   &\left.\left. +2 m_q^4 \sqrt{m_a^4-4 m_\chi^2 m_a^2} \left(-8 m_q m_\chi^4-m_\chi^2 m_a^2
   (m_q+5 m_\chi)+m_a^4 (5 m_\chi-3 m_q)\right) \log \left(\frac{\sqrt{m_a^4-4 m_\chi^2 m_a^2}+m_a^2}{2 m_\chi m_a}\right)\right.\right.\nonumber\\
   &\left.\left. +2 m_q^2
   m_\chi^2 m_a^2 (m_q-m_\chi) \left(m_a^2 \left(-3 m_q^2+2 m_q m_\chi-3 m_\chi^2\right)+8 m_q^2 m_\chi^2\right)\right. \right. \nonumber\\
   &\left.\left.+2 m_\chi^4
   \sqrt{m_a^4-4 m_q^2 m_a^2} \left(8 m_q^4 m_\chi+m_q^2 m_a^2 (5 m_q+m_\chi)+m_a^4 (3 m_\chi-5 m_q)\right) \log
   \left(\frac{\sqrt{m_a^4-4 m_q^2 m_a^2}+m_a^2}{2 m_q m_a}\right)\right.\right.\nonumber\\
   &\left.\left.+m_\chi^4 m_a^4 \left(15 m_q^3-5 m_q^2 m_\chi-5 m_q
   m_a^2+3 m_\chi m_a^2\right) \log \left(\frac{m_q^2}{m_a^2}\right)\right)\right]\;.
\end{align}
}

\bibliographystyle{JHEPfixed}
\bibliography{pseudobib}

\end{document}